\documentclass[12pt]{article}

\usepackage{arxiv}

\usepackage[utf8]{inputenc} 
\usepackage[T1]{fontenc}    
\usepackage{hyperref}       
\usepackage{url}            
\usepackage{booktabs}       
\usepackage{amsfonts}       
\usepackage{nicefrac}       
\usepackage{microtype}      
\usepackage{lipsum}		
\usepackage{graphicx}
\usepackage{doi}
\usepackage{scalerel,stackengine}
\usepackage{graphicx,multicol}
\usepackage{multirow} 
\usepackage{indentfirst,csquotes}
 \usepackage{array} 
\usepackage{abstract}

\usepackage[caption=false]{subfig}
\usepackage{amssymb,amsthm,amsmath}
\usepackage{xcolor,paralist,hyperref,fancyhdr,etoolbox}
\newtheorem{theorem}{Theorem}[]
\newtheorem{definition}[theorem]{Definition}

\newtheorem{lemma}[]{Lemma}

\usepackage[round]{natbib} 

\usepackage[utf8]{inputenc}
\numberwithin{equation}{section}

\usepackage{hyperref}

\hypersetup{ colorlinks=true, linkcolor=black, filecolor=black, urlcolor=black,citecolor = blue }

\usepackage{lipsum}
\usepackage{multicol}

\usepackage{listings}
\title{A Semi-Parametric Torus-to-Torus Regression Model with Geometric Loss: Application to Cyclone Data}

\author{ { Surojit Biswas}\thanks{ webpage: https://sites.google.com/view/surojitbiswas/home?authuser=0} \\
	Department of Mathematics\\  IIT Kharagpur, India-$721302$ \\
	\texttt{surojit23@iitkgp.ac.in} \\
	\And
    {Buddhananda Banerjee}\thanks{ webpage: https://sites.google.com/site/buddhanandastat/} \\
	Department of Mathematics\\  IIT Kharagpur, India-$721302$ \\
	\texttt{bbanerjee@maths.iitkgp.ac.in } \\
}

\date{}

\renewcommand{\shorttitle}{Torus-to-Torus Regression in Cyclone Data}

\hypersetup{
pdftitle={A template for the arxiv style},
pdfsubject={q-bio.NC, q-bio.QM},
pdfauthor={David S.~Hippocampus, Elias D.~Striatum},
pdfkeywords={First keyword, Second keyword, More},
}

\begin{document}
\maketitle
\begin{abstract}

This study introduces a novel torus-to-torus regression framework to improve the analysis and prediction of cyclone-driven wind-wave directional dynamics.
This research, to our knowledge, establishes a mathematical framework for modeling the regression between bivariate angular predictors and bivariate angular responses for the first time in the literature.  The proposed approach enhances the capacity to model coupled directional processes commonly observed in extreme coastal cyclones.
The proposed model makes use of generalized M\"{o}bius transformation and differential geometry for model building.
A new loss function, derived from the intrinsic geometry of the torus, is introduced to facilitate effective semi-parametric  estimation without requiring any specific distributional assumptions on the angular error. 
The prediction error is measured as an angular loss on the surface of the torus and also the angular deflection along normal directions on the unit sphere transported from the torus. 
Additionally, a new visualization technique for circular data is introduced. The practical relevance of the model is illustrated through its application to wind-wave directional datasets from two major cyclonic events, \textit{Amphan} and  \textit{Biparjoy}, that impacted the eastern and western coastlines of India, respectively.
\end{abstract}


\section{Introduction} 

Coastal cyclones are considered one of the most severe natural hazards affecting the Indian subcontinent. The Bay of Bengal, with its warm sea surface temperatures and conducive atmospheric conditions, generates a disproportionately large share of intense cyclonic systems that make landfall along the densely populated eastern coastline. On the contrary, the Arabian Sea has historically produced fewer cyclones, although recent observations indicate an increasing frequency and intensity linked to changing oceanic and atmospheric patterns. These cyclones bring destructive winds, storm surges, and heavy precipitation, resulting in coastal flooding, widespread infrastructure damage, and significant socio-economic disruption. Rapid urbanization in low-lying coastal zones further amplifies vulnerability. Understanding the directional behavior of wind-wave interactions during such events is therefore critical for improving forecasting systems, coastal calamity risk assessment, and climate resilience planning in India’s cyclone-prone coastal regions.

\noindent
\textbf{Challenges with Cyclone Data:}
The interrelation between wave direction and wind direction during a cyclone is dynamic and intricate.  Initially, waves synchronize with dominant winds as strong gusts convey energy to the ocean surface.  As the cyclone progresses, spinning wind patterns result in ongoing changes in wind direction surrounding the eye.  This produces waves that initially align with the wind but then propagate outward autonomously.  The cyclone's trajectory, dimensions, and coastal topography additionally affect wave dynamics.  As waves propagate from the storm center, they are influenced by supplementary environmental elements, including ocean currents and atmospheric conditions.
 The wind-wave interactions of Super Cyclone  \textit{Biparjoy}, which impacted the western coast of India, and \textit{Amphan}, which severely affected the eastern coast of India, present a compelling case study for such modeling methodologies.
However, traditional analysis techniques designed for linear data are not directly suitable for such toroidal (bivariate angular data) or directional datasets. Linear statistical methods implicitly assume that the underlying sample space is unbounded and Euclidean, where distances and averages are computed on a straight line. In contrast, directional variables such as wind and wave directions are defined on a circular $(0^\circ-360^\circ)$ or toroidal domain, where values wrap around. This wrapping causes issues such as misleading mean estimates (e.g., the average of $359^\circ$ and $1^\circ$ should be $0^\circ$, not $180^\circ$) and incorrect distance measurements, leading to artificial discontinuities in the data. Furthermore, correlation structures for directional variables are inherently periodic, violating assumptions of linear correlation models. Consequently, specialized techniques from circular and spherical statistics are required to correctly capture the true central tendency, dispersion, and dependence structure of directional climate variables, particularly under rapidly evolving cyclone dynamics.\\

\noindent
\textbf{Data of  Biparjoy cyclone:} 
\label{Biparjoy section}
The Extremely Severe Cyclonic Storm \textit{Biparjoy} was a long-lived and powerful system that affected parts of western India from June 6 to June 19, 2023. According to the India Meteorological Department (IMD) and the Regional Specialized Meteorological Center - Tropical Cyclones, New Delhi, India Meteorological Department (IMD) (see, \url{https://mausamjournal.imd.gov.in/index.php/MAUSAM/article/view/7221}),  \textit{Biparjoy} was the longest-duration cyclone in the Arabian Sea since 1977.
The storm originated from an upper air cyclonic circulation over the southeast Arabian Sea, which evolved into a depression on June 6. Moving northwards, it intensified into a deep depression and then a cyclonic storm, named \textit{Biparjoy}, in the adjoining region. The system continued to strengthen rapidly, evolving into a Severe Cyclonic Storm (SCS), then a Very Severe Cyclonic Storm (VSCS), and eventually reached the Extremely Severe Cyclonic Storm (ESCS) category by June 11 over the east-central Arabian Sea.
Between June 7 and June 11,  \textit{Biparjoy} followed a recurving path-shifting from north-northwest to north-northeast and then returning to a predominantly northward movement while maintaining ESCS intensity. The cyclone later tracked north-northwest and weakened into a VSCS over the northeast and adjoining east-central Arabian Sea. As it continued its journey, it moved northward and then northeastward, gradually losing intensity.
On June 15,  \textit{Biparjoy} made landfall between Mandvi (Gujarat) and Karachi (Pakistan), near latitude $23.28^{\circ}N$ and longitude $68.56^{\circ}E$, crossing the Saurashtra and Kutch coasts. After landfall, it weakened into a cyclonic storm over Saurashtra and Kutch, and then into a deep depression over southeast Pakistan and adjoining southwest Rajasthan. It continued east-northeastwards, weakening into a depression over south Rajasthan and north Gujarat, and finally into a well-marked low-pressure area over central northeast Rajasthan by June 19.\\
\noindent
\textbf{ Data of  Amphan cyclone:}
The Super Cyclonic Storm (SuCS) \textit{Amphan} caused extensive damage to eastern India and Bangladesh during the period 16th May 2020 to 21st May 2020  over the Bay of Bengal (BoB). As per the description by the Regional Specialized Meteorological Center-Tropical Cyclones, New Delhi, India Meteorological Department (see,  \url{https://internal.imd.gov.in/press_release/20200614_pr_840.pdf}),  \textit{Amphan} started forming from the remnant of a low-pressure area that occurred over the south Andaman Sea and adjoining southeast Bay of Bengal during the period  6th-12th May 2020.  This gradually developed into a well-marked low-pressure area that was observed over southeast BoB at 0300 UTC on 14th May 2020. It intensified into a cyclonic storm over the southeast BoB on the  16th of May 2020. This further intensified into a SuCS over the next two and a half days and maintained the intensity of SuCS for nearly 24 hours before weakening into an Extremely Severe Cyclonic Storm over west-central BoB on 19th May 2020 and making landfall $21.65 ^\circ N, 88.3^\circ E$ on the coast of West Bengal, India, during 1000-1200  UTC on 20th May, with a maximum sustained wind speed of 155 - 165 kmph gusting to 185 kmph.  \\

\noindent
\textbf{Motivation:} The wind-wave directional data are modeled as toroidal data (bivariate angular data). The study of angular data stands apart from conventional linear data due to their inherent periodicity. This characteristic introduces unique analytical challenges, as traditional statistical methods may lead to misleading interpretations when applied directly to circular or toroidal variables. Consequently, specialized statistical frameworks are required to accurately model the directional dependence and variability associated with wind-wave interactions during cyclonic events. Other than wind and wave directions in meteorology \cite[][]{biswas2024angular},  such data commonly emerge across diverse fields, including the assessment of astigmatism angles in ophthalmology \cite[see][]{jha2018circular,banerjee2026intrinsic}, the timing of stock price highs and lows in finance \cite[see][]{biswas2025semi}, agency and communion dimensions in educational psychology \cite[][]{cremers2020regression}, and protein folding angles in bioinformatics \cite[][]{boomsma2008generative}. For some recent developments on directional data, see the book by \cite{kumardirectional}.
A particularly fascinating area within directional statistics is the analysis of bivariate angular data, wherein two angular variables are simultaneously dealt with. Unlike linear or circular data, it ends up belonging to a torus
embedded in $\mathbb{R}^3$. Such a data structure is relevant when two periodic variables are jointly observed. For example, in finance, the times at which a stock attains its high and low values during a day can be represented as toroidal variables.  In environmental studies, wind and wave directions, which are circular and often interdependent, seamlessly align with this paradigm.
 Numerous statistical methods have been developed to address bivariate angular data.  The bivariate von Mises sine model proposed by  \cite{singh2002probabilistic} and the bivariate von Mises cosine model introduced by \cite{mardia2007protein} are potential toroidal distributions. These models elucidate the dependency structure among paired angular variables, providing a foundation for inference for such a data structure.

In a significant contribution, \cite{mccullagh1996mobius} presented the application of M\"{o}bius transformations in directional statistics to establish a link between the Cauchy distribution (in $\mathbb{R}$) and the wrapped Cauchy distribution (on $\mathbb{S}$, unit circle).  A M\"{o}bius transformation, often referred to as a linear fractional transformation in the discipline of complex analysis, is a function that maps the extended complex plane $\mathbb{C} \cup \{\infty\}$  onto itself, while preserving the structure of lines and circles.  This transformation is defined as follows:
\begin{eqnarray}
f(z) = \frac{az + b}{cz + d}
\label{gen_mob_map}
\end{eqnarray}
where $z$ represents a complex variable and $a, b, c, d \in \mathbb{C}$ are constants with the condition that $ad - bc \neq 0$.  These transformations are conformal, signifying that they preserve angles and map circles and lines to other circles or lines.
 Building on this foundation, \cite{jha2018circular} utilized M\"{o}bius transformations as link functions in circular-circular regression, specifically to model zero-inflated angular data frequently seen in ophthalmology.  Previous to that, M\"{o}bius transformations were examined in regression settings by multiple authors, including \cite{downs2002circular}, \cite{downs2003spherical}, and \cite{kato2008circular}, who established diverse regression frameworks for circular and spherical data.

Torus-to-torus regression is a statistical technique intended for examining the correlation between a bivariate circular response variable and a bivariate circular predictor variable.  In contrast to conventional bivariate linear regression, which works inside Euclidean space, this framework accommodates the periodic and topological characteristics of angular data.  The topology of the torus presents distinct issues and requires specialized techniques.  Torus-to-torus regression offers a viable framework for elucidating complex relations between pairs of angular variables, yielding insights that may not be attainable using traditional linear models.
\noindent
\textbf{Our Contributions:}
\begin{itemize}
    \item  We propose a novel torus-to-torus regression model using generalized  M\"{o}bius transformation.

\item  The proposed  M\"{o}bius transformation is a suitable link function because it preserves the toroidal structure and it is free from the identifiability problem.
\item  A new two-component loss function is introduced using concepts from differential geometry. The first component represents the square-angle loss on the surface of the torus. The second component represents square-angle-deflection along normal directions on the unit sphere transported from the torus. The loss function is minimized for the parameter estimation. 

\item  The model is flexible because it is semi-parametric, as it does not assume any specific distribution for the angular error.

\item  A new visualization technique for circular data is introduced for a better understanding of predicted and observed data.

\item  We demonstrate the practical effectiveness of the proposed model through its application to wind and wave direction data recorded during two major cyclones,  \textit{Amphan} and  \textit{Biparjoy}, which impacted the eastern and western coasts of India, respectively. Specifically, we model the prediction of wind-wave direction at 12:00 p.m. based on 6:00 a.m. data for  \textit{Biparjoy}, and at 1:00 p.m. based on 7:00 a.m. data for \textit{Amphan}.

\end{itemize}

This paper is organized as follows.  Section \ref{int_geo_tor} revisits the intrinsic geometry of the torus through foundational tools from differential geometry.  Section \ref{regression model sec} illustrates the usefulness of the M\"{o}bius transformations for circular-circular and linear-circular regressions. 
Section \ref{tor_reg_model_sec} presents a comprehensive development of the torus-to-torus regression model.  Section \ref{tor_reg_curve} provides the construction of the regression curve and examines its identifiability issue in parameter space. 
Section \ref{tor_reg_model} incorporates the non-parametric assumptions about the angular error and analyzes its geometric properties. 
Section \ref{loss fun sec} introduces a novel loss function derived from the intrinsic geometry of the torus, facilitating effective parameter estimation. 
Section \ref{simulation sec} provides a comprehensive analysis of simulation studies conducted across diverse angular error distributions to evaluate model performance.  A novel visualization technique, the \textit{Circular Scatter Plot}, is presented in Section \ref{csctr_plot} as a diagnostic instrument for assessing directional model fit.  The model is utilized in Section \ref{data_analysis} on real wind and wave direction data gathered during the super cyclones  \textit{Biparjoy} and \textit{Amphan}.  Concluding remarks are presented in Section \ref{conclusion}.

\section{Intrinsic Geometry of Torus}
\label{int_geo_tor}
Here, a new area-based measurement between two angles $\phi, \theta \in \Omega$ is briefly discussed. The rest of our work will be based on the curved torus defined by the parametric equation 
\begin{equation}
  X(\phi,\theta)=\{  (R+r\cos{\theta})\cos{\phi}, (R+r\cos{\theta})\sin{\phi}, r\sin{\theta} \}\subset \mathbb{R}^3, 
  \label{torus para equn}
\end{equation}
With the parameter space $\{ 
 (\phi,\theta):0<\phi,\theta<2\pi\}= \Omega \times \Omega,$ known as  flat torus and $R>r$. 
Taking  partial derivatives of $X$ with respect to $\phi$, and $\theta$ are 
$$ X_{\phi}=\{ -(R+r\cos{\theta})\sin{\phi}, (R+r\cos{\theta})\cos{\phi}, 0 \}$$ and
$$X_{\theta}=\{-r\sin{\theta} \cos{\phi}, -r\sin{\theta}\sin{\phi}, r\cos{\theta}  \}$$ respectively. Hence, the coefficients of the first fundamental form are
 \begin{equation}
 \begin{aligned}
     E=\langle X_{\phi},X_{\phi}\rangle &= (R+r\cos{\theta})^2\\
    F=\langle X_{\phi},X_{\theta}\rangle &= 0 \\
     G=\langle X_{\theta},X_{\theta}\rangle &= r^2
 \end{aligned}
 \label{fff cof}
 \end{equation}
leading to the area element of the curved torus (Equation-\ref{torus para equn}) 
\begin{equation}
    dA=\sqrt{EG-F^2} ~d\phi~d\theta=r(R+r\cos{\theta})~d\phi~d\theta
    \label{torus area element}
\end{equation}

The following cross product gives the unit normal vector on the surface of the curved torus:
\begin{eqnarray}
   {N} = \frac{{X}_\phi}{\sqrt{E}} \times \frac{{X}_\theta}{\sqrt{G}} =(\cos\phi \cos\theta, \sin\phi \cos\theta,
\sin\theta).
\label{tor_norml}
\end{eqnarray}

 Now using this area element \cite{biswas2025semi} have introduced, ``square of an angle'' denoted by $A_C^{(0)}(\theta)$, which is the minimum area between $(0,0)$ to  $(\theta, \theta)$ on the curved torus. The formal definition is as follows:

\begin{definition}

     \textit{The square of an angle} $\theta$ is defined as
   $$A_C^{(0)}(\theta)=A_T\left[(0,0),(\theta,\theta)\right]=\frac{\min \{A_1,A_2,A_3,A_4\}}{4\pi^2rR}.
    \label{def:sq_of_angle}
$$
\end{definition}
Here  $A_1, A_2, A_3,$ and $A_4$ denote the four distinct surface areas on the curved torus, calculated as $A_i = \iint_{\mathbb{T}_i} dA(s,t)$ for $i = 1,2,3,4$, where $dA(s,t)$ is the  area element as defined in Equation-\ref{torus area element}. The regions $\mathbb{T}_1 = [0, \theta] \times [0, \theta]$, $\mathbb{T}_2 = [\theta, 2\pi] \times [0, \theta]$, $\mathbb{T}_3 = [0, \theta] \times [\theta, 2\pi]$, and $\mathbb{T}_4 = (\theta, 2\pi] \times [\theta, 2\pi]$ constitute a partition of $(0, 2\pi] \times (0, 2\pi]$.
The notion of the square of an angle has been generalized to the surface of a sphere and a torus by \cite{biswas2024angular} as follows:
\begin{equation}
    A_C^{(0)}(\theta)=\displaystyle 
    \begin{cases}
        A_T\left[(0,0),(\theta,\theta)\right]=A_{T}^{(0)}(\theta)& \text{for the curved torus}\\
        A_S\left[(0,0),(\theta,\theta)\right] =A_{S}^{(0)} (\theta)& \text{for the sphere.}
    \end{cases}
    \label{eq: square_ang_tor_sph}
\end{equation}

We use these new area-based measurements, $A_T^{(0)}(\theta)$ and $A_S^{(0)}(\theta)$ to construct the least-squares equivalent condition for the torus-to-torus regression.



\section{ M\"{o}bius Transformation-Based Regression Models in the Literature} \label{regression model sec} 
\textbf{Circular-circular regression:}
The typical linear regression framework can represent interactions between circular variables, as demonstrated by \cite{hussin2004linear}.  This method exhibits significant constraints within the circular-circular framework.  The model specifically fails to produce consistent predictions for inputs that differ by complete rotations (i.e., $x$ and $x + 2\pi$) unless the regression coefficient is an integer.  Furthermore, the method demonstrates susceptibility to the arbitrary selection of the zero-angle reference, leading to non-equivalent regression functions across varying orientations of the independent circular variable.  These challenges highlight the insufficiency of traditional linear regression for circular-circular data and necessitate the study of alternative modeling approaches that preserve the topological and geometric properties of the circle.  One effective method is to employ the circle-to-circle M\"{o}bius transformation. In the complex plane $\mathbb{C}$, this map is a closed group transformation. For the set 
$\Omega=\{ z: z\in \mathbb{C}, |z|=1\}$, the circle to circle M\"{o}bius map is defined by 
\begin{eqnarray}
    f(z)=e^{i\alpha}\frac{z+a}{1+\bar{a}z}
    \label{cir_mob_map}
\end{eqnarray}
$\alpha\in (0,2\pi]$, $a\in \mathbb{C}, 
\bar{a}$ is conjugate of $a,$ and $|a| \neq 1.$

Recently, \cite{jha2018circular, jha2017multiple} utilized this transformation as a link function for circular-circular and multiple circular-circular regression, respectively. They also provided a geometric interpretation of the model parameters (see,Figure-\ref{model_para_interptation}(b)).
Previous applications of M\"{o}bius transformations in circular-circular regression models are documented in the studies by \cite{kato2008circular} and \cite{downs2002circular}. All the aforementioned studies utilize a parametric framework. For instance, \cite{jha2018circular, jha2017multiple} employed the von Mises distribution \cite[see][p.36]{mardia2000directional}, while \cite{kato2008circular} utilized a wrapped Cauchy distribution \cite[see][p.51]{mardia2000directional} to model angular error.

\noindent
\textbf{Linear-circular regression:}
The M\"{o}bius transformation that maps the upper half of the complex plane $\mathbb{C}$ (Im$(z) > 0$) onto the open unit disk ${w \in \mathbb{C}: |w| < 1}$, and its boundary (Im$(z)=0$) onto the unit circle ($|w| = 1$), is given by
\begin{equation}
w = M(z; \beta_0, \beta_1) = \beta_0 \frac{z - \beta_1}{z - \overline{\beta_1}},
\label{full model}
\end{equation}
where $\beta_0 \in \Omega = {w: |w| = 1}$ acts as a rotation parameter and $\beta_1 \in \mathbb{C}$ defines the center of the transformation by mapping to the origin in the $w$-plane \cite[see][Sec. 95]{brown2009complex}. In linear-circular regression, the predictor $x \in \mathbb{R}$ and the response $y$ lie on the unit circle. Adapting Equation-\ref{full model}, \cite{biswas2025semi} proposed a semi-parametric  regression model:
\begin{equation}
y = \beta_0 \frac{x - \beta_1}{x - \overline{\beta_1}} \epsilon,
\label{full real model}
\end{equation}
where $\arg(\epsilon)$ is a zero-mean angular error. The model is semi-parametric , making no distributional assumptions on $\epsilon$, and parameter estimation is achieved via a novel area-based loss function derived from the intrinsic geometry of the torus.
Another application of the M\"{o}bius transformation in the linear-circular regression model can be seen in \cite{kim2015inverse}. In this case, they have used a particular M\"{o}bius map. Also, this is a parametric approach where the angular error follows von Mises distribution and asymmetric generalized von Mises distribution.

\section{The Proposed torus-to-torus  Regression Model}
\label{tor_reg_model_sec}
Let $(u_i, v_i)$, $i=1,\cdots, n$ be the response variables corresponding to the nonstochastic covariates, $(z_i, w_i)$ for $i=1,\cdots, n$. The observations are independent of each other. Both, $(u, v)$  and $(z, w)$ take values on the torus, $\mathbb{T}_2=\Omega \times \Omega$. The proposed torus-to-torus regression model is semi-parametric ; hence, it is applicable for any  conditional distribution of $(u, v)$ given $(z,w)$ maintaing some basic assumptions of the random angular error. 
In the section below, we introduce the regression curve and examine its theoretical properties.

\subsection{Regression Curve}\label{tor_reg_curve}
Let $\eta \in \mathbb{R}\setminus \{0,1\}$, $\beta_0, \gamma_0 \in \Omega$ and $\beta_1, \gamma_1 \in \mathbb{C}$  be complex parameters with the conditions that $|\beta_1| \neq 1$ and $|\gamma_1| \neq 1$.  The regression curve for the proposed model is characterized by 
 \begin{eqnarray} 
(f_1(z,w; \eta, \beta_0, \beta_1), f_2(z,w; \eta, \gamma_0, \gamma_1)) 
= \left( \beta_0\frac{z^\eta+w\beta_1}{w+\overline{\beta_1}z^\eta}, \gamma_0 \frac{w+z^\eta\gamma_1 }{z^\eta+\overline{\gamma_1}w} \right), \label{tor_mob_map} 
 \end{eqnarray}
 where $z,w \in \Omega$. Both components of the proposed map are M\"{o}bius maps.  For instance, let us examine the first component and set $\beta_0=1$ then $f_1(z,w; \eta, \beta_0, \beta_1)=\frac{z^\eta+w\beta_1}{w+\overline{\beta_1}z^\eta}$.  Upon comparison with the general M\"{o}bius map presented in Equation-\ref{gen_mob_map}, it is evident that $a=1, b=w\beta_1, c=\bar{\beta}_1$, and $d=w.$  In the case of the general M\"{o}bius map, the parameters $a$, $b$, $c$, and $d$ are all complex constants.  However, in the proposed model, $d=w$ is  a complex variable rather than a constant.  The same reasoning applies to the second component of the proposed model.

In the proposed torus-to-torus regression model, the parameter $ \eta $ serves as an \textit{interaction parameter} that captures nonlinear dependencies between the two circular predictors within the M\"{o}bius transformation framework. Given that the predictors $ z $ and $ w $ lie on the unit circle in the complex plane, raising $ z $ to the power $ \eta $ corresponds to a rescaling of its angular component; that is, if $ z = e^{i\theta} $, then $ z^\eta = e^{i \eta \theta} $. To ensure unambiguous interpretation, we adopt the convention that for $ z \in \Omega = \{ e^{i\theta} : \theta \in [0,2\pi) \} $ and real $ \eta $, the transformation is defined as $ z^\eta = e^{i \eta \theta} $, where $ \theta $ is the principal argument of $ z $. Equivalently, if $ z = \cos\theta + i\sin\theta $, then $ z^\eta = \cos(\eta\theta) + i\sin(\eta\theta) $. This definition guarantees that $ z^\eta $ is single-valued and continuous on $ [0,2\pi) $, thereby aligning $ \eta $ with a continuous rescaling of the angular coordinate. Importantly, the values $ \eta = 0 $ and $ \eta = 1 $ are excluded, since $ \eta = 0 $ eliminates any dependence on a predictor, reducing the model to a degenerate form, while $ \eta = 1 $ imposes a restrictive linear-like relationship that limits the capacity of the model to adequately capture the nonlinear interactions intrinsic to toroidal data. Consequently, we define the admissible parameter space as $ \eta \in \mathbb{R} \setminus \{0,1\} $, a restriction that preserves both the expressiveness and interpretability of the regression structure.  

\noindent
\textbf{Branch‐Cut Limitations and Numerical Robustness:}\label{sec:branchcut}
The torus‐to‐torus map defined in \eqref{tor_mob_map} involves non‐integer powers \(z^\eta\) for \(z\in\Omega\).  Since $z^\eta = \exp\bigl(i\,\eta\,\arg(z)\bigr),$
the complex logarithm introduces a branch cut discontinuity along the negative part of the real axis. As the estimator is a continuous-valued random variable, the estimator falling in the branch cut has probability zero. Additionally, to ensure the branch cut issue does not compromise model validity, we perform the following numerical diagnostic tests:
\begin{enumerate}[]
  \item \textbf{Residual discontinuity and prediction continuity check:}  Plot the circular residuals 
    \(\delta_\phi =[(\phi_\mathrm{pred}-\phi_\mathrm{obs}) \pmod{2\pi} ]\)
    against \(\arg(z)\) and \(\arg(w)\).  Large spikes near \(-\pi\) or \(\pi\) indicating branch‐cut effects.
  \item \textbf{Cut‐placement sensitivity:}  Rotate all predictor angles by \(30^\circ\) and refit.  We record the percentage change in each parameter estimate.  Changes under 5\% are deemed robust.
  \item \textbf{Multiple starts:}  Run the optimizer from multiple random initial points.  
\end{enumerate}
Results of these tests are presented in Section~\ref{sec:branch_cut}, demonstrating that the parameter estimation and predictions remain stable despite the theoretical branch‐cut limitation.  Now, each component of the proposed map in Equation-\ref{tor_mob_map} can be decomposed with the 
fundamental types of maps:
\begin{itemize}

   \item \textbf{Translation:} Shift \(x\) by \(b\): $x \rightarrow x +b.$

    \item \textbf{Inversion:} Apply an inversion: $x \rightarrow  \frac{1}{x}.$

    \item \textbf{Scaling:} Scale by \(\alpha\): $x \rightarrow  \alpha x.$
\end{itemize}
For example, the first component from Equation-\ref{tor_mob_map} can be decomposed, and
$$f_1(z,w; \eta, \beta_0, \beta_1)=\beta_0\left( \frac{1}{\overline{\beta_1}}+\frac{\tau}{\overline{\beta_1}z^\eta+w}\right),~~~~~ \mbox{where}~~~\tau=w\beta_1-\frac{w}{\overline{\beta_1}}.$$  
A similar decomposition is possible for $f_2(z,w; \eta, \gamma_0, \gamma_1) $.

\begin{lemma}
    The map defined in Equation-\ref{tor_mob_map} maps a torus to a torus.
\end{lemma}
\begin{proof}
    In order to prove it, we need to show that \\
    $f_1(z,w; \eta, \beta_0, \beta_1) \in \Omega, f_2(z,w; \eta, \gamma_0, \gamma_1)\in \Omega $.
    Let us consider 
    \begin{eqnarray}
       | f_1(z,w; \eta, \beta_0, \beta_1)|^2&=&\left| \beta_0\frac{z^\eta+w\beta_1}{w+\overline{\beta_1}z^\eta} \right|^2 \nonumber\\
       &=&\frac{\left|z^\eta+w\beta_1\right|^2}{|w+\overline{\beta_1}z^\eta|^2} \nonumber\\
       &=&\frac{(z^\eta+w\beta_1)\overline{(z^\eta+w\beta_1)}}{(w+\overline{\beta_1}z^\eta)\overline{(w+\overline{\beta_1}z^\eta)}}\\
       && [~~\text{since~~}|\beta_0|=1, |z^\eta|=|w|=1]\nonumber\\
       &=&\frac{1+z^\eta\bar{w}\overline{\beta_1}+w\bar{z^\eta}\beta_1+|\beta_1|^2}{1+z^\eta\bar{w}\overline{\beta_1}+w\bar{z^\eta}\beta_1+|\beta_1|^2} \nonumber\\
       \implies | f_1(z,w; \eta, \beta_0, \beta_1)|&=&1 \nonumber\\
       \implies  f_1(z,w; \eta, \beta_0, \beta_1) &\in& \Omega. \nonumber
    \end{eqnarray}
Similarly, it can be demonstrated for \( f_2(z,w; \eta, \gamma_0, \gamma_1) \). 
 Therefore, the lemma is established.
\end{proof}



\subsubsection{Identifiability of the Model}  The identifiability of models for the multiple circular-circular regression has been extensively examined by \cite{jha2017multiple}.  Here we address the same for the proposed torus-to-torus regression.
Consider a point \((z, w) \in \mathbb{T}_2\) and a parameter \((\beta_1, \gamma_1) \in \mathbb{C} \times \mathbb{C}\). Setting \(\beta_0 = \gamma_0 = 1\), the predicted point \((u, v)\) on the torus \(\mathbb{T}_2\) is obtained via the M\"{o}bius transformation (Equation~\ref{tor_mob_map}). Geometrically, \((u, v)\) lies along the line in \(\mathbb{C} \times \mathbb{C}\) connecting \((-z, -w)\) and \((\beta_1, \gamma_1)\). Therefore, any alternate parameter \((\beta_1', \gamma_1')\) yielding the same predicted point for the same \((z, w)\) must also lie on this line. Since this condition holds for all \((z, w) \in \mathbb{T}_2\), the points \((-z, -w)\), \((\beta_1, \gamma_1)\), and \((\beta_1', \gamma_1')\) must be collinear.
Now, if we select a point, 
$(z', w') \ne \left(\beta_0 \frac{z^\eta + w\beta_1}{w + \overline{\beta_1}z^\eta}, \gamma_0 \frac{w + z^\eta\gamma_1}{z^\eta + \overline{\gamma_1}w}\right),$
collinearity of the three points \((-z', -w')\), \((\beta_1, \gamma_1)\), and \((\beta_1', \gamma_1')\) holds if and only if \((\beta_1, \gamma_1) = (\beta_1', \gamma_1')\). Hence, the parameters \((\beta_1, \gamma_1)\) uniquely determine the model. More formally, we can state that in the following theorem.

\begin{theorem}
\label{thm:identifiability_precise2}
Let 
$\boldsymbol\xi_1=(\eta,\beta_0,\beta_1),
\boldsymbol\xi'_1=(\eta',\beta_0',\beta_1'), \quad
\boldsymbol\xi_2=(\eta,\gamma_0,\gamma_1),
\boldsymbol\xi'_2=(\eta',\gamma_0',\gamma_1'),$ with
$\eta,\eta'\in\mathbb R\setminus\{0,1\},~~~~~~
\beta_0,\gamma_0,\beta_0',\gamma_0'\in\Omega,~~~~~
\beta_1,\gamma_1,\beta_1',\gamma_1'\in\mathbb C,$
and assume that
$|\beta_1| \neq 1, |\gamma_1|\neq 1, \quad 
|\beta_1'|\neq1, |\gamma_1'|\neq1.$
If $[f_1(z,w;\boldsymbol\xi_1), f_2(z,w;\boldsymbol\xi_2)]=[f_1(z,w;\boldsymbol\xi'_1), f_2(z,w;\boldsymbol\xi'_2)]$ for all $(z,w)\in\Omega^2$,\\ then
$\boldsymbol\xi_1=\boldsymbol\xi'_1  ;~\boldsymbol\xi_2=\boldsymbol\xi'_2.$
\end{theorem}
\begin{proof}
Consider the first component. Suppose
\begin{equation}\label{eq:comp1}
\beta_0\frac{z^\eta+w\beta_1}{w+\overline{\beta_1}z^\eta}
=\beta_0'\frac{z^{\eta'}+w\beta_1'}{w+\overline{\beta_1'}z^{\eta'}}
\quad\text{for all }(z,w)\in\Omega^2.
\end{equation}
Cross-multiplying denominators yields
\begin{align*}
\beta_0\big(z^\eta + w\beta_1\big)\big(w+\overline{\beta_1'}z^{\eta'}\big)
&=\beta_0'\big(z^{\eta'}+w\beta_1'\big)\big(w+\overline{\beta_1}z^\eta\big)\\
\iff
z^\eta w(\beta_0-\beta_0'\beta_1'\overline{\beta_1})
&+z^{\eta'}w(\beta_0\beta_1\overline{\beta_1'}-\beta_0') \\
&+z^{\eta+\eta'}(\beta_0\overline{\beta_1'}-\beta_0'\overline{\beta_1}) +w^2(\beta_0\beta_1-\beta_0'\beta_1')=0.
\end{align*}

If \(\eta\neq\eta'\), the four monomials
\[
z^\eta w,\quad z^{\eta'}w,\quad z^{\eta+\eta'},\quad w^2
\]
have distinct exponent pairs and are linearly independent as functions on \(\Omega^2\). Hence their coefficients must vanish separately, giving
\[
\beta_0=\beta_0'\beta_1'\overline{\beta_1},\quad
\beta_0\beta_1\overline{\beta_1'}=\beta_0',\quad
\beta_0\overline{\beta_1'}=\beta_0'\overline{\beta_1},\quad
\beta_0\beta_1=\beta_0'\beta_1'.
\]
From the second and fourth equalities,
\(\beta_0'=\beta_0\beta_1\overline{\beta_1'}=\beta_0\beta_1/\beta_1'\),
so \(\overline{\beta_1'}=1/\beta_1'\), implying \(|\beta_1'|=1\).  
This contradicts the hypothesis \(|\beta_1'|\neq1\).  
Therefore \(\eta=\eta'\).

With \(\eta=\eta'\), set \(t=w/z^\eta\). Then \eqref{eq:comp1} becomes
\[
\beta_0\frac{1+t\beta_1}{t+\overline{\beta_1}}
=\beta_0'\frac{1+t\beta_1'}{t+\overline{\beta_1'}}.
\]
Both sides are M\"{o}bius transformations in \(t\), and they agree for infinitely many \(t\in\Omega\). Hence they coincide identically, which implies
\[
\beta_1=\beta_1',\qquad \beta_0=\beta_0'.
\]

Applying the same reasoning to the second component gives
\(\gamma_1=\gamma_1'\) and \(\gamma_0=\gamma_0'\). Therefore, \(\boldsymbol\xi_i=\boldsymbol\xi_i'\) for \(i=1,2\).  
\end{proof}

\begin{figure*}[h!]
	\centering
\includegraphics[trim= 0 0 0 0, clip, width=0.55\textwidth,height=.35\textwidth]{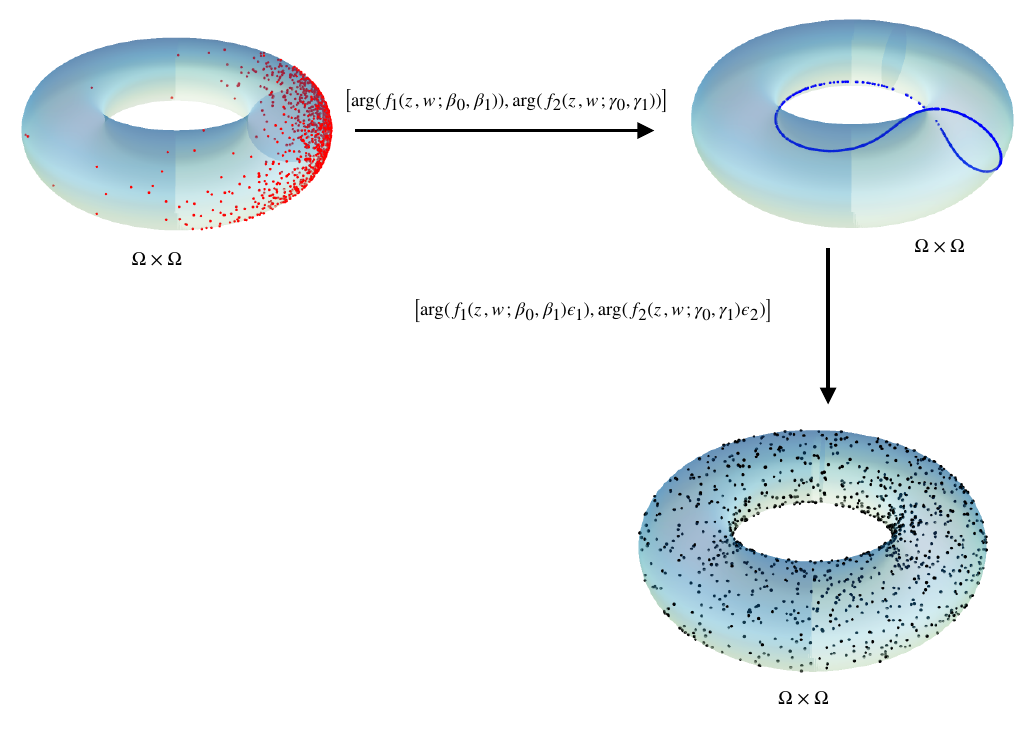}
	\caption{Top-left: scatter plots depicting the predictor values on the torus $\mathbb{T}_2$.  
 Top-right: Graph illustrating the expected responses via the proposed M\"{o}bius map on the torus $\mathbb{T}_2$.   Bottom-right: A plot illustrating the responses with angular error, $\arg\boldsymbol{\epsilon}=(\arg \epsilon_1, \arg \epsilon_2)$, derived from a von Mises sine model with zero mean direction.}
    \label{fig: model_plot}
\end{figure*}
\subsection{The Regression Model} \label{tor_reg_model}
Using the analogy of multiple linear regression in Euclidean spaces, where covariates can collectively influence the response, we define the toroidal regression model, using Equation-\ref{tor_mob_map}, as follows:
\begin{eqnarray}
( u, v)&=& \left[\beta_0\frac{z^\eta+w\beta_1}{w+\overline{\beta_1}z^\eta}\epsilon_1,  \gamma_0 \frac{w+z^\eta\gamma_1 }{z^\eta+\overline{\gamma_1}w}\epsilon_2 \right] \mbox{~~which implise} \nonumber \\ 
   (\phi,\theta)&=& \left[\arg\left( \beta_0\frac{z^\eta+w\beta_1}{w+\overline{\beta_1}z^\eta}\epsilon_1\right), \arg\left( \gamma_0 \frac{w+z^\eta\gamma_1 }{z^\eta+\overline{\gamma_1}w}\epsilon_2 \right)\right]\nonumber \\ 
   &=& \begin{bmatrix}
\{\arg (f_1(z,w; \eta, \beta_0, \beta_1)) + \arg (\epsilon_1)\}\pmod{2\pi}  \\
\{\arg (f_2(z,w; \eta, \gamma_0, \gamma_1)) + \arg (\epsilon_2) \}\pmod{2\pi}
\end{bmatrix}^t\nonumber\\ 
    &\equiv&  \left(\begin{bmatrix}
g_1(z,w; \eta, \beta_0, \beta_1)\\
g_2(z,w; \eta, \gamma_0, \gamma_1)
\end{bmatrix}^t +\begin{bmatrix}
 \arg(\epsilon_1)  \\
\arg (\epsilon_2)
\end{bmatrix}^t \right)\pmod{2\pi} 
    \label{tor_reg_model_full}
\end{eqnarray}
In this context, $g_1$ and $g_2$ represent link functions, while $\phi, \theta \in [0, 2\pi)$. The parameters $\beta_0$, $\gamma_0$, $\epsilon_1$, and $\epsilon_2 \in \Omega$, and $\beta_1$ and $\gamma_1$ are complex constants, with the conditions that $|\beta_1| \neq 1$ and $|\gamma_1| \neq 1$.  The vector $ \arg (\boldsymbol{\epsilon}) = (\arg (\epsilon_1), \arg (\epsilon_2))$ denotes a random angular error with a zero mean direction.  In alignment with the classical multiple linear regression framework, where the error term $\boldsymbol{\epsilon}$ is assumed to have a zero mean (e.g., $ \boldsymbol{Y} =  \mathbf{X}\boldsymbol{\beta} + \boldsymbol{\epsilon}$), we posit that the angular error in the proposed model similarly exhibits a zero mean direction and fixed concentration parameters.  This assumption is reasonable from a modeling standpoint. 
 Upon obtaining estimates for $\beta_0, \gamma_0$ and $\beta_1, \gamma_1$, the corresponding regression curve can be derived using Equation-\ref{tor_mob_map}.  A pictorial depiction of the
construction of the model is shown in Figure-\ref{fig: model_plot}.

It is important to emphasize that the angular error does not follow any particular distribution and has zero mean direction and constant concentrations across observations. Consequently, the only parameters requiring estimation are those associated with the M\"{o}bius transformation. This modeling framework thus constitutes a \textit{semi-parametric  torus-to-torus regression model}, combining the flexibility of nonparametric error modeling with a parametric structure for the regression function.

\subsubsection{Geometry of the Proposed Model}
\begin{figure*}[h!]
    \centering
    \subfloat[]{%
        {\includegraphics[trim= 10 10 10 10, clip,width=0.33\textwidth, height=0.25\textwidth]{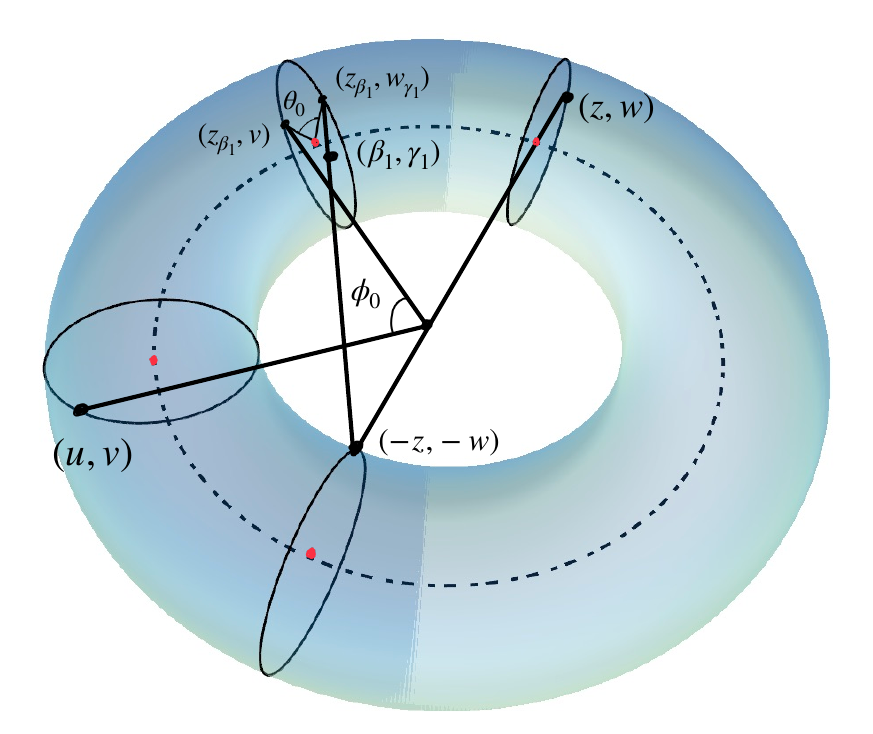}}}\hspace{15pt}
    \subfloat[ ]{%
        {\includegraphics[trim= 0 0 0 0, clip,width=0.23\textwidth, height=0.23\textwidth]{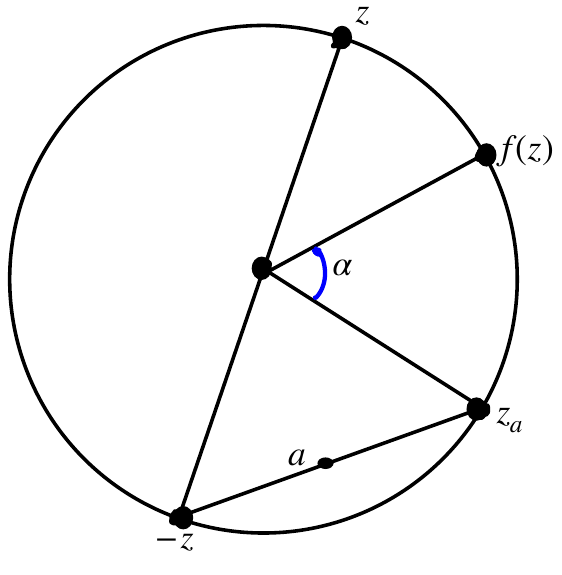}}}
    \caption{ (a) torus-to-torus regression model. (b) Circular-circular regression model.}
     \label{model_para_interptation}
\end{figure*}

The proposed model presented in Equation-\ref{tor_mob_map} utilizes a total of four complex parameters ($\beta_0, \gamma_0, \beta_1, \gamma_1$) and one real parameter ($\eta$).  For simplicity let us consider $\eta=1$, $\beta_0, \gamma_0 \in \Omega$ and $\beta_1, \gamma_1 \in \mathbb{C}$, subject to the conditions $|\beta_1| \neq 1$ and $|\gamma_1| \neq 1$.  It is evident from the setup that $\beta_0$ and $\gamma_0$ operate as rotation parameters.  The interpretation of $\beta_1$ and $\gamma_1$ is complicated.  The interpretation, as illustrated in Figure-\ref{model_para_interptation}(a), can be described as follows.  Consider the pair $(z,w)$ located on the positive curvature region of the torus, denoted as $\Omega \times \Omega$.  We now rotate the point by $\pi$ radians horizontally and $\pi$ radians vertically.  Thus, we obtain the point $(-z,-w)$, which represents the antipodal point on the torus situated in the region of negative curvature ( A similar approach can be seen in circular-circular regression due to \cite{jha2018circular}).   This antipodal point $(-z,-w)$ intersects the line that passes by $(\beta_1, \gamma_1)$ and meets the torus at the coordinates, $(z_{\beta_1},w_{\gamma_1})$. We now fix $z_{\beta_1}$ and rotate $w_{\gamma_1}$ by $\theta_0 = \arg(\gamma_0)$, resulting in an intermediate point, $(z_{\beta_1},v)$. Next, fix $v$ and rotate $z_{\beta_1}$ by $\phi_0 = \arg(\beta_0)$, yielding the final regressed point $(u, v)$. In this way,  under the combined effect of the four complex parameters $\beta_0, \gamma_0 \in \Omega$ and $\beta_1, \gamma_1 \in \mathbb{C}$, the point $(z,w)$ is transformed to the regressed point $(u,v)$.






\noindent
\textbf{Note:}
\begin{itemize}

    \item Maintaining one component in a fixed position while rotating the other, as previously described, is not necessary.  Instead, the point $(z_{\beta_1}, w_{\gamma_1})$ can be simultaneously rotated using the angles $(\phi_0, \theta_0)$ to directly obtain the regressed point $(u, v)$.  The step-by-step explanation is for better understanding, as illustrated in Figure-\ref{model_para_interptation}(a).
    \item The above geometric interpretation is for a model where we choose the covariate $(z,w)$ and the response $(u,v)$. Now, since $z\in\Omega$ then, for any interaction  parameter $\eta\in \mathbb{R}\setminus \{0,1\}$, $z^\eta\in \Omega$. Hence, the geometric interpretation above is valid for the proposed model in Equation-\ref{tor_mob_map} as well.
\end{itemize}

\subsubsection{Rotational Equivariance and Invariance of Parameters}
Toroidal regression models must exhibit invariance under rotations, meaning that the behavior of the model should not be influenced by the arbitrary selection of an origin on the torus.  Thus, when the covariates and responses undergo rotation by specified angles, it is essential for the predictions to remain the same, assuming that the model parameters are suitably modified. 
The map in Equation-\ref{tor_mob_map} can be written as 
\begin{eqnarray}
    (u,v)&=& \bigg\{ \beta_0  \exp\left[ i \arg \left( \frac{z^\eta+w\beta_1}{w+\overline{\beta_1}z^\eta}\right)\right],  \gamma_0 \exp\left[ i \arg \left(\frac{w+z^\eta\gamma_1 }{z^\eta+\overline{\gamma_1}w}\right)\right]\bigg\}  \nonumber\\ 
    &=& \bigg\{ \beta_0  \exp\left[ i \arg \left( f((z, w; \eta , \beta_1)\right)\right], \gamma_0 \exp\left[ i \arg \left(f(z, w; \eta, \gamma_1)\right)\right]\bigg\},
    \label{tor_mob_map2}
\end{eqnarray}
where $z,w \in \Omega$, $\beta_0, \gamma_0$ are rotation parameters.  We will now examine the following scenario:

\noindent
\textbf{Rotation of the responses and covariates:}
A modification in the reference direction, specifically at the point $(0,0)$ on the torus, corresponds to the multiplication of unit complex numbers $W_1$ and $W_2$, where $|W_1|= |W_2|= 1$.  From Equation-\ref{tor_mob_map2}, it is evident that if the response variable $(u,v)$ is transformed to $(W_1u,W_2v)$, then the model which yield identical predictions is given by $\big\{ W_1\beta_0  \exp\left[ i \arg \left( f((z, w; \eta , \beta_1)\right)\right],  W_2\gamma_0 \exp\left[ i \arg \left(f(z, w;\eta, \gamma_1)\right)\right]\big\}.$

When the covariates,  $(z, w)$  is rotated by $(W_1, W_2)$, then the model that yield the same predictions as the original one by $\bigg\{ \bar{W_1}W_2\beta_0 ~ \exp\left[ i \arg \left( f(z, w;\frac{W_1}{W_2}\beta_1)\right)\right],  
\bar{W_2}W_1\gamma_0~ \exp\left[ i \arg \left(f(z, w;\frac{W_2}{W_1}\gamma_1)\right)\right]\bigg\},$ after the rotations.




\begin{figure*}[h!]
    \centering
    \subfloat[]{%
        {\includegraphics[trim= 90 90 80 90, clip,width=0.32\textwidth, height=0.27\textwidth]{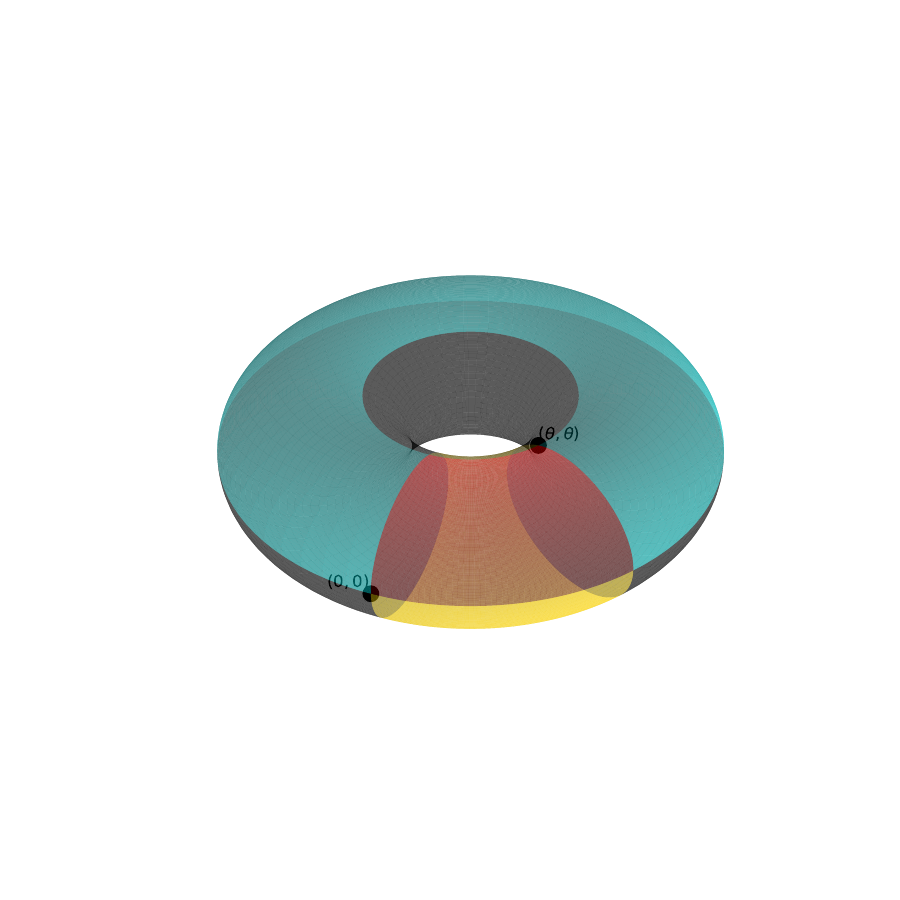}}}
    \subfloat[ ]{%
        {\includegraphics[trim= 0 0 0 0, clip,width=0.32\textwidth, height=0.22\textwidth]{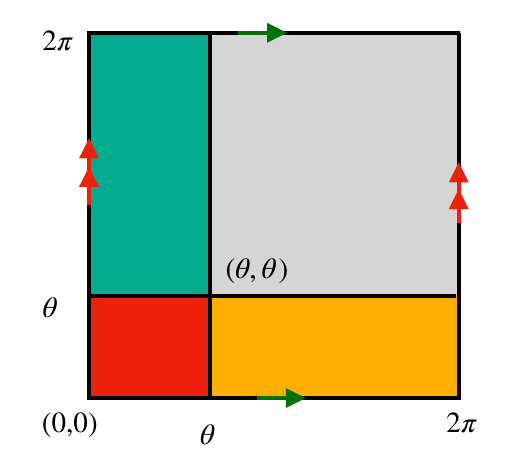}}}
                \subfloat[]{%
        {\includegraphics[trim= 0 0 0 0, clip,width=0.4\textwidth, height=0.25\textwidth]{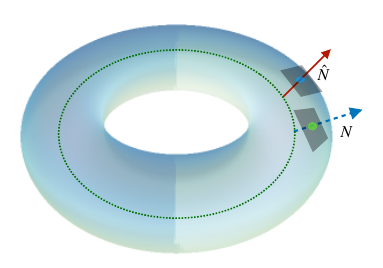}}}
        
    \subfloat[ ]{%
        {\includegraphics[trim= 0 0 0 0, clip,width=0.3\textwidth, height=0.28\textwidth]{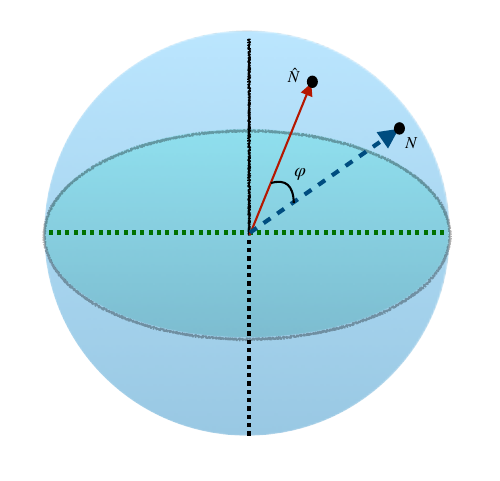}}}
        \subfloat[]{%
        {\includegraphics[trim= 90 135 90 90, clip,width=0.36\textwidth, height=0.34\textwidth]{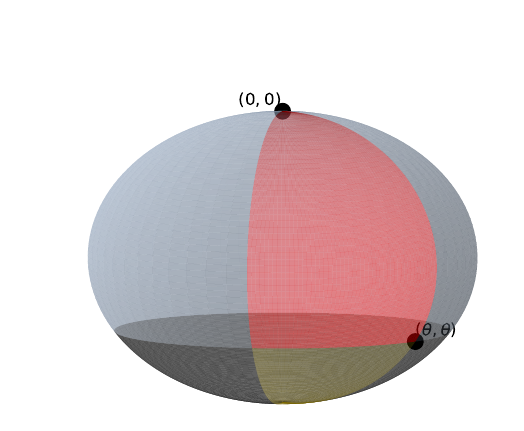}}}
    \subfloat[ ]{%
        {\includegraphics[trim= 0 0 0 0, clip,width=0.34\textwidth, height=0.24\textwidth]{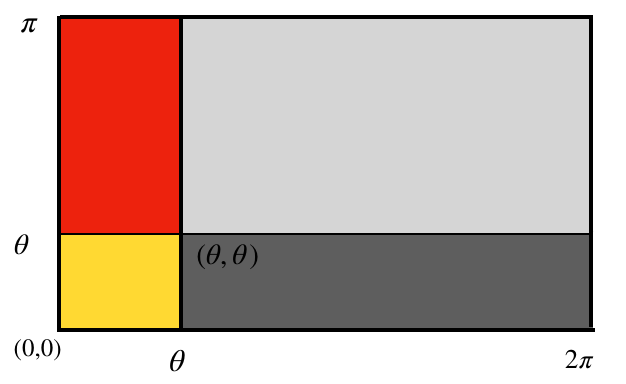}}}
        \hspace{3pt}
    \caption{ (a) Area between $(0,0)$, and $(\theta, \theta)$ on the curved torus. (b) Area between $(0,0)$, and $(\theta, \theta)$ in a flat torus. (c) The normals at observed and predicted points. (d) Geodesic distance between two normals on a sphere. (e) Area between $(0,0)$, and $(\theta, \theta)$ on sphere. (f) Area between $(0,0)$, and $(\theta, \theta)$ in a flat sphere.}
     \label{confidence interval}
\end{figure*}

\section{Loss Function and  Estimation} \label{loss fun sec}
In least-squares regression analysis, from $\mathbb{R}^2$ to $\mathbb{R}^2$, the model is defined as

$$
y_i = A X_i + b + \varepsilon_i, \quad i = 1, \dots, n.
$$
where
$$
X_i =
\begin{bmatrix} X_{i1} \\ X_{i2} \end{bmatrix}
\in \mathbb{R}^2, \quad
y_i =
\begin{bmatrix} y_{i1} \\ y_{i2} \end{bmatrix}
\in \mathbb{R}^2,
A =
\begin{bmatrix} a_{11} & a_{12} \\ a_{21} & a_{22} \end{bmatrix}
\in \mathbb{R}^{2 \times 2},
$$
and $b =
\begin{bmatrix} b_1 \\ b_2 \end{bmatrix}
\in \mathbb{R}^2,$ $ \epsilon_{i} =
\begin{bmatrix} \epsilon_{i1}  \\ \epsilon_{i2}  \end{bmatrix}$ is the zero-mean random error vector.
The component-wise mean squared error (MSE) is employed as the loss function to quantify the average squared difference between the predicted values and the actual observed values. By minimizing the MSE, we estimate the model parameters that provide the best fit to the data. This minimization ensures that the predicted values are as close as possible to the observed ones, reducing the overall error in the regression model. Now, for observed values \( y_i \) and predicted values \( \hat{y}_i \) (\( i = 1, 2, \cdots, n \)), the MSE can be computed as:
\[
\text{MSE} =\frac{1}{n}
 \sum_{i=1}^{n} \left[ (y_{i1} - \hat{y}_{i1})^2 + (y_{i2} - \hat{y}_{i2})^2 \right]=\frac{1}{n}
 \sum_{i=1}^{n} \left[ |\mathbb{Y}_{i1}|^2 + |\mathbb{Y}_{i2}|^2 \right],
\]
where each \( |\mathbb{Y}_{ij}|^2 = |\left(y_{ij} - \hat{y}_{ij}\right)|^2 \), for \( i = 1, 2, \cdots, n \)  and $j=1,2$ geometrically represents the mean of random squared areas with arm length $|\mathbb{Y}_{ij}|$ in \( \mathbb{R}^2 \) for \(i = 1, 2, \cdots, n \) and $j=1,2$. 
Following the definition of  MSE and its geometric interpretation, \cite{biswas2025semi} introduced the \textit{mean square-angle error} \textit{(MSAE)} using Definition-\ref{def:sq_of_angle} as 
MSAE = $ \displaystyle\frac{1}{n}\sum_{i=1}^{n} A_{0}^{C}(a_i),
    \label{loss function}$ where, $a_i$ being any angular difference.

\subsection{Proposed Loss Function} 
Let $[(u_i, v_i), (z_i, w_i)] \in \mathbb{T}_2 \times \mathbb{T}_2$ such that $|u_i|=|v_i|=|z_i|=|w_i|=1$ for \(i=1,2,\cdots,n\) be the observed data points. Using Equation-\ref{tor_reg_model_full}, the expected values for $(u_i, v_i)$ a given $(z_i, w_i)$ can be written as
\[
E[(\phi_i, \theta_i)|(z_i, w_i)]=  [ g_1(z,w; \eta, \beta_0, \beta_1) , g_2(z,w; \eta, \gamma_0, \gamma_1)]
\]
 for \(i=1,2,\cdots,n\).
The  angular difference between observed and expected values is defined as  
\begin{eqnarray}
    (\psi_i, \xi_i )&=& \left[(\phi_i, \theta_i) -E\{(\phi_i, \theta_i)|(z_i, w_i)\} \right] \pmod{2\pi}\nonumber\\ 
    &=& \left[(\phi_i, \theta_i) - g_1(z,w; \eta, \beta_0, \beta_1) , g_2(z,w; \eta, \gamma_0, \gamma_1) \right] \pmod{2\pi}, \nonumber\\
    &=& \Bigl[ \{\phi_i - g_1(z,w; \eta, \beta_0, \beta_1)\} \pmod{2\pi}, \nonumber \\
             &&\{\theta_i - g_2(z,w; \eta, \gamma_0, \gamma_1)\} \pmod{2\pi} \Bigr] , ~~\text{for~} i=1,\cdots,n.\nonumber
\end{eqnarray}
We utilize the geometric interpretation of \textit{mean square-angle error} \textit{(MSAE)} to propose one of the components of the loss function based on the `square of an angle' defined for  the curved torus in Equation-\ref{eq: square_ang_tor_sph} as:
\begin{eqnarray}
    \mathcal{L}_T(\beta_0, \gamma_0, \beta_1, \gamma_1)= \frac{1}{n}\sum_{i=1}^{n} [A_{T}^{(0)}(\psi_i)+A_{T}^{(0)}(\xi_i)].
    \label{loss function torus}
\end{eqnarray}

The proposed regression model is based on the torus, $\mathbb{T}_2$.  In contrast to the flat torus, $[0, 2\pi) \times [0, 2\pi)$, it encompasses all three types of curvature: zero, positive, and negative.  To achieve more effective optimization of the model parameters, we incorporate another component to the loss function based on the shortest distance (geodesic or great circle distance) between the normals at the observed and predicted values on the torus.

        
        

It is evident from Equation-\ref{tor_norml} that each normal represents a point on a unit sphere defined by the following parametric equation.
\begin{eqnarray}
    Z(p, t)=(\cos p \cos t, \sin p \cos t,
\sin t), ~\text{where}~~ p \in [0,2\pi], t\in[0,\pi].
\label{sphere_para}
\end{eqnarray}
Now, let $(\phi_i, \theta_i), ( \hat{\phi_i}, \hat{\theta_i})  \in \mathbb{T}_2$ for $i=1,2, \cdots,n$ be the observed and estimated (through the model in Equation-\ref{tor_reg_model_full}) points, respectively. Then the unit normals corresponding to the points are 
\begin{eqnarray}
    N_i &=&(\cos\phi_i \cos\theta_i, \sin\phi_i \cos\theta_i,
\sin\theta_i) ;~~~\mbox{and} \\
 \hat{N}_i&=&(\cos \hat{\phi_i} \cos \hat{\theta_i}, \sin \hat{\phi_i} \cos\hat{\theta_i},
\sin\hat{\theta_i}), \nonumber
\end{eqnarray}
respectively for $i=1,2, \cdots,n$. Hence, the distance between two unit normals $N_i, \hat{N_i}$ is the shortest distance (the great circle distance) on the sphere, which is  given by 
\begin{eqnarray}
    \varphi_i=\cos^{-1}[\sin \theta_i \sin\hat{\theta_i}+ \cos \theta_i \cos\hat{\theta_i} \cos (\phi_i-\hat{\phi_i})],
\end{eqnarray}
$\text{for}~~~ i=1,2, \cdots,n.$ Because of the unit radius, the angle between $N_i$ and $\hat{N}_i$ is $\varphi_i$. In this instance, similar to the first component of the loss function,  we compute another component of the loss function for the normals again using  Equation-\ref{eq: square_ang_tor_sph} for sphere,  as follows: 
\begin{eqnarray}
     \mathcal{L}_S(\beta_0, \gamma_0, \beta_1, \gamma_1)=  \frac{1}{n}\sum_{i=1}^{n} [A_{S}^{(0)}(\varphi_i)].
    \label{loss function_sphere}
\end{eqnarray}

Hence, combining the Equation-\ref{loss function torus} and Equation-\ref{loss function_sphere}, we define the loss function to estimate the parameters of the proposed regression model as follows:
\begin{eqnarray}
   \mathcal{L}(\beta_0, \gamma_0, \beta_1, \gamma_1)&=&\mathcal{L}_T(\beta_0, \gamma_0, \beta_1, \gamma_1)+ \mathcal{L}_S(\beta_0, \gamma_0, \beta_1, \gamma_1)\nonumber\\
   &=&  \frac{1}{n}\sum_{i=1}^{n} \bigg\{ [A_{T}^{(0)}(\psi_i)+ A_{T}^{(0)}(\xi_i)]+ [A_{S}^{(0)}(\varphi_i)] \bigg\}.
    \label{loss function total}
\end{eqnarray}
The estimated values of the parameters $\beta_0, \gamma_0, \beta_1, \gamma_1$ can be calculated by minimizing $\mathcal{L}(\beta_0, \gamma_0, \beta_1, \gamma_1)$.
 Due to the lack of a closed-form solution for the parameters in the loss function, various numerical optimization techniques may be utilized to minimize the loss function defined in Equation-\(\ref{loss function total}\) and get the estimated values of $(\beta_0, \gamma_0, \beta_1, \gamma_1)$ as (\(\hat{\beta}_0\), \(\hat{\gamma}_0\), \(\hat{\beta}_1\), \(\hat{\gamma}_1\)).  This investigation employed a numerical optimization method, which is particularly effective for problems with bounds.  This method  facilitates an effective parameter estimation by minimizing the loss function for the proposed model.\\

\section{ Simulation} \label{simulation sec}
This section presents a comprehensive simulation analysis to assess the efficacy of the proposed torus-to-torus regression model.  The study analyzes the model with varied parameter specifications over three distinct sample sizes \((n)\) to evaluate its robustness and performance.  We examine \(n = 50\) to investigate small-sample behavior, \(n = 150\) for moderate sample size impacts, and \(n = 500\) to analyze large-sample performance.
To illustrate that the angular error in the proposed regression model is independent of the distribution function, we utilize two separate angular error distributions, the von Mises sine model, and the von Mises cosine model. The von Mises sine model was introduced by \cite{singh2002probabilistic} with the probability density function
\begin{eqnarray}
    f_{vMsine}(\phi,\theta)&=&\frac{1}{C} \exp \Bigl[ \kappa_1\cos(\phi-\mu_{\phi})+\kappa_2\cos(\theta-\mu_{\theta}) + \kappa_3 \sin(\phi-\mu_{\phi})\sin(\theta-\mu_{\theta}) \Bigr]
    \label{sine model bv}
\end{eqnarray}
where , $(\mu_{\phi},\mu_{\theta})\in [0, 2\pi)$, $\kappa_1, \kappa_2>0$, $\kappa_3 \in \mathbb{R}$, and $C$, the normalizing constant, is given by 
$$C=4\pi^2\displaystyle \sum_{m=0}^{\infty}  \binom{2m}{m} \bigg(  \frac{\kappa_3^2}{4\kappa_1\kappa_2} \bigg)^m I_m(\kappa_1)I_m(\kappa_2),$$  and
$I_m(\kappa)$ denotes the modified Bessel function of the first kind of order $m.$ The von Mises cosine model was proposed by \cite{mardia2007protein}  with the probability density function 
\begin{eqnarray}
    f_{vMcos}(\phi,\theta)&=&\frac{1}{C} \exp \Bigl[\{\rho_1\cos(\phi-\mu_{\phi})+\rho_2\cos(\theta-\mu_{\theta}) +\rho_3 \cos(\phi-\mu_{\phi}-\theta+\mu_{\theta})    \Bigr]
             \label{cosine model bv}
\end{eqnarray}

where, $(\mu_{\phi},\mu_{\theta})\in [0, 2\pi)$, $\rho_1, \rho_2>0$, $\rho_3 \in \mathbb{R}$, and $C$, the normalizing constant, is given by 
$$C=4\pi^2\displaystyle  \bigg[ I_0(\rho_1)I_0(\rho_2)I_0(\rho_3)+ 2 \displaystyle \sum_{m=0}^{\infty}  I_m(\rho_1)I_m(\rho_2)I_m(\rho_3)\bigg],$$  and
$I_m(\rho)$ denotes the modified Bessel function of the first kind of order $m.$  

Simulations are conducted for both distributions with a fixed mean direction vector, \(\boldsymbol{\mu}=(\mu_1,\mu_2) = (0,0)\) (in radians), fixed concentration parameters, and altering dependency parameters.  This extensive configuration enables us to examine the performance of the model  under various distributional assumptions and sample size conditions.  The simulation outcomes for the angular error models are summarized across several scenarios. For computational simplicity, we express the parameter \(\beta_0\) as \(\exp(i \phi_0)\), while \(\beta_1\) is represented as \(b_1 + i b_2\).   Similarly, \(\gamma_0\) is represented as \(\exp(i \theta_0)\), while \(\gamma_1\) is expressed as \(b_3 + i b_4\).
 The angular errors for the outcome summarized in Table-\ref{table:von_sine_table} are derived from a von Mises sine model with the parameters $(\phi_0, b_1, b_2, b_3, b_4, \theta_0) = (1.0472, -1.7, 1.2, -1.8, 1.5, 3.1416)$.  
Table-\ref{table:von_cos_model_table} presents the simulation  results for angular errors derived from a von Mises cosine model with parameters $(\phi_0, b_1, b_2, b_3, b_4, \theta_0) = (0, 0.3, -3.5, 1.7, 0.5, 0)$. For both cases, we conduct 1000 iterations to obtain parameter estimates and their empirical standard errors for sample sizes of 50, 100, and 500, with dependence parameters $\kappa_3 = -1, 0, 1$.  The findings indicate a significant reduction in standard error as sample size increases.  In both instances, covariates are drawn from a von Mises distribution with mean direction $\mu=0,$ and the concentration parameter $\kappa=1$.   The von Mises distribution has the following density function
\begin{equation}
	f_{\text{vm}}(\theta;\mu,\kappa)=\frac{e^{\kappa\cos(\theta-\mu)}}{2\pi I_{0}(\kappa)},
	\label{von mises}
\end{equation}
where $0\leq \theta<2\pi$, $0\leq \mu<2\pi$, $\kappa>0$, and $ I_{0}(\kappa)$ is the modified Bessel function with order zero evaluated at $\kappa.$

\begin{table*}[h!]
\centering
\renewcommand{\arraystretch}{1.2}
\resizebox{1\linewidth}{!}{%
  \begin{tabular}{|c|c|c|c|c|c|c|c|c|}
\hline
& & \multicolumn{7}{c|}{Parameters} \\ \hline
Sample size & Dependency parameter & $\phi_0=\tfrac{\pi}{3}=1.0472$ & $b_1=-1.7$ & $b_2=1.2$ & $b_3=-1.8$ & $b_4=1.5$ & $\theta_0=\pi=3.1416$ & $\eta=2$ \\ \hline

\multicolumn{1}{|c}{\multirow{3}{*}{$n=50$}} &
\multicolumn{1}{|c|}{$\kappa_3=-1$} & 1.0477 (0.0196)& -1.7140 (0.0180)& 1.2066 (0.0169 )& -1.9349 (0.0224)& 1.4276 (0.0200) &2.9887 (0.0142 )& 2.001 (0.018)\\ 
\multicolumn{1}{|c}{} &
\multicolumn{1}{|c|}{$\kappa_3=0$} & 1.0571 (0.0203) &-1.6764 (0.0143)& 1.1992 (0.0150)& -1.9248 (0.0196)& 1.4202 (0.0212) &2.9937 (0.0122) & 1.995 (0.017) \\ 
\multicolumn{1}{|c}{} &
\multicolumn{1}{|c|}{$\kappa_3=1$} & 1.0502 (0.0185) & -1.6687 (0.0148) & 1.1895 (0.0158) & -1.9466 (0.0235) & 1.4644 (0.0216) &3.0050 (0.0120) & 2.008 (0.016) \\ \hline

\multicolumn{1}{|c}{\multirow{3}{*}{$n=150$}} &
\multicolumn{1}{|c|}{$\kappa_3=-1$} & 1.0350 (0.0106)& -1.7262 (0.0106)& 1.1960 (0.0101)& -1.8591 (0.0113)& 1.4538 (0.0112 )& 3.0626 (0.0067 ) & 1.997 (0.012)\\ 
\multicolumn{1}{|c}{} &
\multicolumn{1}{|c|}{$\kappa_3=0$} & 1.0401 (0.0102)& -1.6973 (0.0096)& 1.1906 (0.0092)& -1.8698 (0.0099)& 1.4370 (0.0100)& 3.0547 (0.0073 ) & 2.004 (0.010)  \\ 
\multicolumn{1}{|c}{} &
\multicolumn{1}{|c|}{$\kappa_3=1$} & 1.0444 (0.0108)& -1.6882 (0.0095)& 1.2093 ( 0.0095)& -1.8867 (0.0118) & 1.4547 (0.0108)& 3.0486 (0.0077 ) & 2.010 (0.011) \\ \hline

\multicolumn{1}{|c}{\multirow{3}{*}{$n=500$}} &
\multicolumn{1}{|c|}{$\kappa_3=-1$} & 1.0116 (0.0057)& -1.7086 (0.0056)& 1.1951 (0.0050)& -1.8621 (0.0071)& 1.4676 (0.0063)& 3.0761 (0.0047 ) & 2.002 (0.008)\\ 
\multicolumn{1}{|c}{} &
\multicolumn{1}{|c|}{$\kappa_3=0$} & 1.0406 (0.0059)& -1.7119 (0.0054)& 1.2016 (0.0046)& -1.8514 (0.0059)& 1.5006 ( 0.0057)& 3.0839 (0.0046) & 2.006 (0.007)  \\ 
\multicolumn{1}{|c}{} &
\multicolumn{1}{|c|}{$\kappa_3=1$} & 1.0466 (0.0060)& -1.7319 (0.0054)& 1.2037 (0.0048)& -1.8499 (0.0069)& 1.4946 (0.0049)& 3.0809 (0.0046) & 1.999 (0.007) \\ \hline

\end{tabular}}
\vspace{0.3cm}
\caption{Estimates of the parameter vector (with varying sample sizes) 
$(\phi_0, b_1, b_2, b_3, b_4, \theta_0, \eta) = (1.0472, -1.7, 1.2, -1.8, 1.5, 3.1416, 2)$, 
with standard errors in parentheses. The angular errors are from a von Mises sine model with zero mean direction, fixed concentration parameters $\kappa_1 = \kappa_2 = 3$, and varying dependency parameter $\kappa_3$.}
\label{table:von_sine_table}
\end{table*}

\begin{table*}[h!]
\centering
\renewcommand{\arraystretch}{1.5}
\resizebox{1\linewidth}{!}{%
  \begin{tabular}{|c|c|c|c|c|c|c|c|c|}
\hline
& & \multicolumn{7}{c|}{Parameters} \\ \hline
Sample size & Dependency parameter & $\phi_0=0$ & $b_1=0.3$ & $b_2=-3.5$ & $b_3=1.7$ & $b_4=0.5$ & $\theta_0=0$ & $\eta=0.7$ \\ \hline


\multirow{3}{*}{$n=50$} &
$\rho_3=-1$ &
0.0208 (0.0127) &
0.2498 (0.0209) &
-3.2602 (0.0177) &
1.7149 (0.0054) &
0.5003 (0.0058) &
0.0016 (0.0075) &
0.7019 (0.0075)
\\

&
$\rho_3=0$ &
0.0348 (0.0114) &
0.2341 (0.0188) &
-3.3918 (0.0147) &
1.7070 (0.0045) &
0.5074 (0.0050) &
0.0030 (0.0061) &
0.7005 (0.0061)
\\

&
$\rho_3=1$ &
0.0530 (0.0099) &
0.1845 (0.0178) &
-3.4360 (0.0162) &
1.6978 (0.0045) &
0.4884 (0.0054) &
0.0119 (0.0062) &
0.6992 (0.0062)
\\ \hline


\multirow{3}{*}{$n=150$} &
$\rho_3=-1$ &
-0.0339 (0.0081) &
0.3554 (0.0138) &
-3.3278 (0.0125) &
1.7320 (0.0036) &
0.4898 (0.0039) &
0.0301 (0.0044) &
0.6991 (0.0044)
\\

&
$\rho_3=0$ &
0.0087 (0.0074) &
0.2789 (0.0130) &
-3.4574 (0.0107) &
1.7199 (0.0030) &
0.4871 (0.0028) &
0.0220 (0.0036) &
0.7004 (0.0036)
\\

&
$\rho_3=1$ &
0.0380 (0.0069) &
0.2307 (0.0124) &
-3.5453 (0.0101) &
1.7111 (0.0023) &
0.4970 (0.0027) &
0.0109 (0.0032) &
0.7000 (0.0032)
\\ \hline


\multirow{3}{*}{$n=500$} &
$\rho_3=-1$ &
-0.0393 (0.0048) &
0.3606 (0.0078) &
-3.3439 (0.0076) &
1.7280 (0.0018) &
0.4900 (0.0020) &
0.0256 (0.0025) &
0.7004 (0.0025)
\\

&
$\rho_3=0$ &
0.0023 (0.0040) &
0.2961 (0.0072) &
-3.5120 (0.0070) &
1.7182 (0.0015) &
0.4987 (0.0016) &
0.0116 (0.0019) &
0.6997 (0.0019)
\\

&
$\rho_3=1$ &
0.0312 (0.0036) &
0.2416 (0.0063) &
-3.5747 (0.0065) &
1.7125 (0.0014) &
0.4927 (0.0014) &
0.0136 (0.0017) &
0.7003 (0.0017)
\\ \hline

\end{tabular}}
\vspace{0.3cm}
\caption{
Estimates of the parameter vector 
$(\phi_0, b_1, b_2, b_3, b_4, \theta_0, \eta) = (0, 0.3, -3.5, 1.7, 0.5, 0, 0.7)$
with standard errors in parentheses.  
Angular errors follow the von Mises cosine model with fixed 
$\rho_1=\rho_2=4$ and varying dependency parameter $\rho_3$.}
\label{table:von_cos_model_table}
\end{table*}

To evaluate the robustness of the proposed model, we examine a mixture distribution for angular error, comprising the von Mises sine model and the von Mises cosine model.  For a mixing proportion of $0.5$, the density of the mixture distribution is given by \begin{eqnarray} f_{\text{mix}}(\theta) = \frac{1}{2}\left[f_{vMsine}(\phi,\theta) + f_{vMcos}(\phi,\theta)\right]. \label{mix_dist} \end{eqnarray}
We now consider a more general parameter configuration: $\phi_0 = \frac{\pi}{4} = 0.7854$, $b_1 = -3.3$, $b_2 = -5.5$, $b_3 = -4.7$, $b_4 = -3.1$, and $\theta_0 = \frac{\pi}{4} = 0.7854$. The angular error is generated from a mixture distribution defined in Equation-\ref{mix_dist}, using an equal mixing proportion of $0.5$ for both components. For the von Mises sine component, the concentration parameters are set to $\kappa_1 = 4$ and $\kappa_2 = 5$, while for the von Mises cosine component, the parameters are $\rho_1 = 5$ and $\rho_2 = 6$. The results are summarized in Table-\ref{table:wrap_mix_model_table}, based on a total sample size of 50 and 250. The covariates in this setting are drawn from the wrapped Cauchy distribution, which has the probability density function:
 \begin{equation} f_{\text{wc}} 
 (\theta;\mu,\zeta)=\frac{1}{2\pi}\dfrac{1-\zeta^2}{1+\zeta^2-2\zeta \cos({\theta-\mu})}, \label{wrap cauchy} \end{equation}
 where $0\leq \theta<2\pi$, $0\leq \mu<2\pi$, and $0\leq \zeta <1$.  For the covariates, we have selected $\mu=\pi$ and a concentration parameter of $\zeta=0.2$.   Different combinations of dependency parameters ($\kappa_3\mbox{~and~} \rho_3$) for the angular error are utilized to conduct 1000 iterations, resulting in estimated parameter values and their corresponding empirical standard errors. 
 The analysis indicates that the estimates are satisfactory within this general framework.  Other sample sizes and the combination of dependency parameters related to the angular error may also be examined; however, we omit this discussion for brevity.

\begin{table*}[h!]
\centering
\renewcommand{\arraystretch}{1.5} 
\resizebox{1\linewidth}{!}{%
  \begin{tabular}{|c|c|c|c|c|c|c|c|c|}
\hline
& & \multicolumn{7}{c|}{Parameters} \\ \hline
Sample size & Dependency parameter & $\phi_0=\tfrac{\pi}{4}=0.7854$ & $b_1=-3.3$ & $b_2=-5.5$ & $b_3=-4.7$ & $b_4=-3.1$ & $\theta_0=\tfrac{\pi}{4}=0.7854$ & $\eta=-0.3$ \\ \hline

\multicolumn{1}{|c}{\multirow{3}{*}{$n=50$}} & 
\multicolumn{1}{|c|}{$\kappa_3=\rho_3=0$} & 0.7972 (0.0039) & -3.3509 (0.0065) & -5.5047 (0.0063) & -4.6837 (0.0059) & -3.0951 (0.0053) & 0.7802 (0.0074) & -0.2930 (0.0051) \\  

\multicolumn{1}{|c}{} &
\multicolumn{1}{|c|}{$\kappa_3=-3.35,\rho_3=2.04$} & 0.7839 (0.0065) & -3.3422 (0.0062) & -5.5620  (0.0052) & -4.6901 (0.0061) & -3.0960 (0.0065) & 0.7835 (0.0065) & -0.2996  (0.0055) \\ \hline

\multicolumn{1}{|c}{\multirow{2}{*}{$n=250$}} & 
\multicolumn{1}{|c|}{$\kappa_3=\rho_3=0$} & 0.7843 (0.0018) & -3.2917 (0.0030) & -5.5319 (0.0044) & -4.6924 (0.0044) & -3.0858 (0.0030) & 0.7866 (0.0072) & -0.3004 (0.0014) \\  

\multicolumn{1}{|c}{} &
\multicolumn{1}{|c|}{$\kappa_3=-3.35,\rho_3=2.04$} &  0.7818 (0.0017 ) & -3.3106 (0.0030) & -5.5212 (0.0044) & -4.6698 (0.0044) & -3.0824 (0.0030) & 0.7864 (0.0014 ) & -0.3002 (0.0014) \\ \hline

\end{tabular}}
\vspace{0.3cm}
\caption{Estimates of the parameter vector \((\phi_0, b_1, b_2, b_3, b_4, \theta_0, \eta)=(\tfrac{\pi}{4}, -3.3, -5.5, -4.7, -3.1, \tfrac{\pi}{4}, -0.3)\), with standard errors in parentheses.  The angular error is generated from a mixture distribution defined in Equation-\ref{mix_dist}. The covariates follow $f_{\text{wc}} (\theta;\pi,0.2)$.}
\label{table:wrap_mix_model_table}
\end{table*}

The results in Table-\ref{table:radi_ratio_model_table} illustrate the robustness of the proposed model across various values of the ratio $\frac{r}{R}$ where $r$ and $R$ are the vertical and the horizontal radius of the torus, respectively.  Despite variations in this ratio ranging from 0.1 to 1, the parameter estimates for $\phi_0$, $b_1$, $b_3$, $b_4$, and $\theta_0$ exhibit notable stability and proximity to their true values.  The estimates of the  coefficients $b_1 = 0.3$, $b_2 = 0.5$, $b_3 = 0.7$, and $b_4 = 0.4$ demonstrate minor fluctuations, accompanied by consistently low standard errors.  The angular parameters $\phi_0$ and $\theta_0$ exhibit minor deviations from their target values of 0 and $\pi$, respectively, indicating that the model successfully adjusts to variations in toroidal curvature while preserving estimation accuracy.

\begin{table*}[h!]
\centering
\renewcommand{\arraystretch}{1.2}
\resizebox{0.9\linewidth}{!}{%
  \begin{tabular}{|c|c|c|c|c|c|c|c|c|}
\hline
& & \multicolumn{7}{c|}{Parameters} \\ \hline
Sample size & Ratio of radius & $\phi_0=0$ & $b_1=0.3$ & $b_2=0.5$ & $b_3=0.7$ & $b_4=0.4$ & $\theta_0=\pi=3.1416$ & $\eta=1.4$ \\ \hline

\multicolumn{1}{|c}{\multirow{10}{*}{$n=250$}} &
$\frac{r}{R}=0.1$ & -0.0159 (0.0022) & 0.2937 (0.0014) & 0.5002 (0.0014) & 0.6961 (0.0011) & 0.4061 (0.0010) & 3.1346 (0.0010) & 1.4006 (0.0010) \\

& $\frac{r}{R}=0.2$ & -0.0286 (0.0028) & 0.2878 (0.0015) & 0.5016 (0.0016) & 0.6943 (0.0011) & 0.4066 (0.0011) & 3.1343 (0.0010) & 1.4009 (0.0010) \\

& $\frac{r}{R}=0.3$ & -0.0308 (0.0029) & 0.2911 (0.0016) & 0.5036 (0.0017) & 0.6928 (0.0012) & 0.4085 (0.0012) & 3.1363 (0.0009) & 1.3997 (0.0009) \\

& $\frac{r}{R}=0.4$ & -0.0354 (0.0028) & 0.2881 (0.0016) & 0.5034 (0.0019) & 0.6921 (0.0013) & 0.4093 (0.0013) & 3.1361 (0.0009) & 1.3994 (0.0009) \\

& $\frac{r}{R}=0.5$ & -0.0312 (0.0028) & 0.2868 (0.0016) & 0.4973 (0.0018) & 0.6931 (0.0012) & 0.4073 (0.0014) & 3.1343 (0.0011) & 1.4001 (0.0011) \\

& $\frac{r}{R}=0.6$ & -0.0328 (0.0028) & 0.2850 (0.0017) & 0.4961 (0.0020) & 0.6917 (0.0013) & 0.4092 (0.0013) & 3.1363 (0.0009) & 1.4014 (0.0009) \\

& $\frac{r}{R}=0.7$ & -0.0391 (0.0029) & 0.2859 (0.0017) & 0.5015 (0.0018) & 0.6938 (0.0013) & 0.4081 (0.0013) & 3.1366 (0.0008) & 1.4008 (0.0008) \\

& $\frac{r}{R}=0.8$ & -0.0344 (0.0031) & 0.2841 (0.0019) & 0.4985 (0.0018) & 0.6959 (0.0015) & 0.4088 (0.0015) & 3.1350 (0.0012) & 1.3993 (0.0012) \\

& $\frac{r}{R}=0.9$ & -0.0324 (0.0029) & 0.2873 (0.0017) & 0.4996 (0.0019) & 0.6925 (0.0013) & 0.4099 (0.0016) & 3.1331 (0.0012) & 1.4003 (0.0012) \\

& $\frac{r}{R}=1$ & -0.0360 (0.0028) & 0.2872 (0.0019) & 0.5002 (0.0019) & 0.6924 (0.0013) & 0.4093 (0.0016) & 3.1355 (0.0011) & 1.3994 (0.0011) \\ \hline

\end{tabular}}
\vspace{0.3cm}
\caption{Estimates of the parameter vector $(\phi_0, b_1, b_2, b_3, b_4, \theta_0, \eta) =(0, 0.3, 0.5, 0.7, 0.4, \pi, 1.4)$ are presented, with standard errors in parentheses, for varying ratios of vertical and horizontal radius of the torus. The sample size is $n=250$.}
\label{table:radi_ratio_model_table}
\end{table*}

\begin{figure*}[h!]
    \centering
    \subfloat[]{%
        {\includegraphics[trim= 2 2 2 2, clip,width=0.25\textwidth, height=0.25\textwidth]{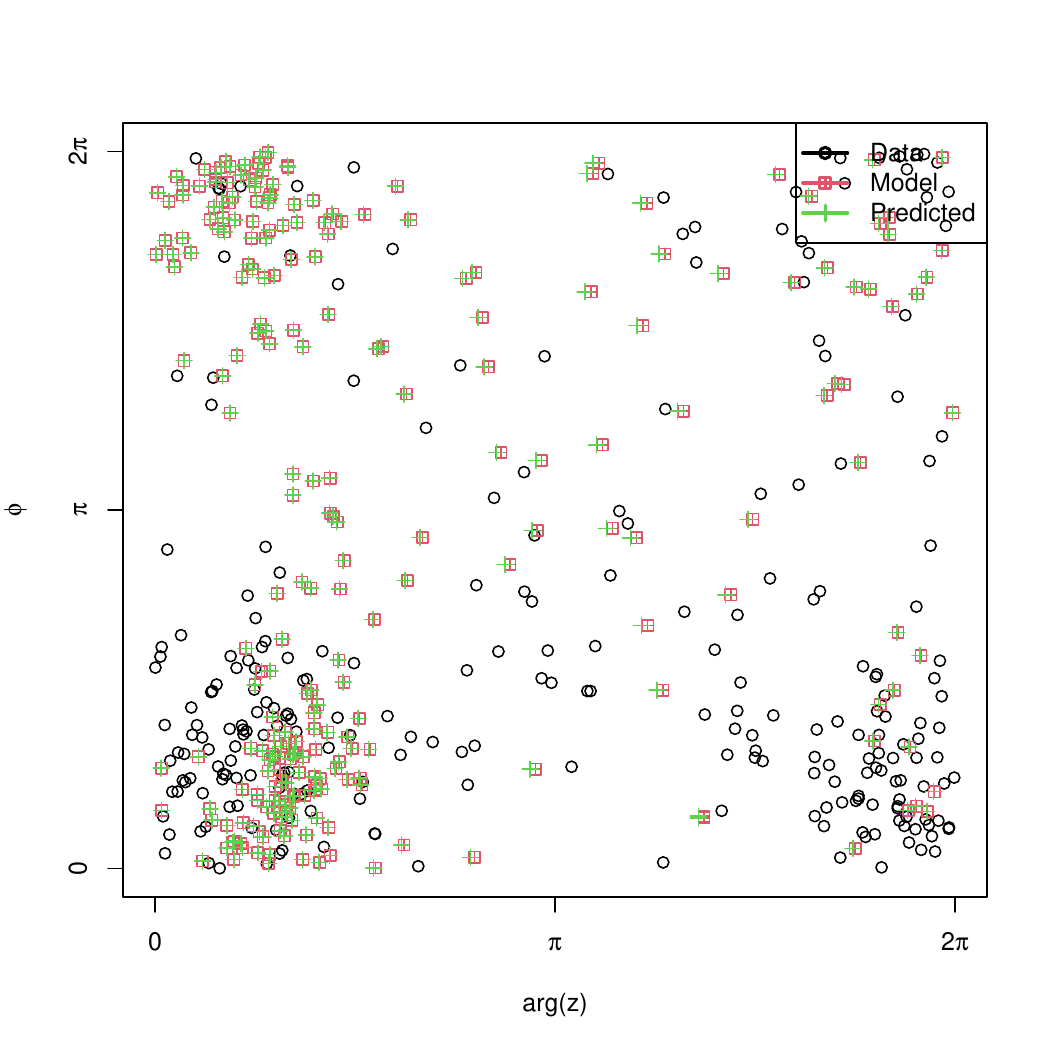}}}\hspace{5pt}
    \subfloat[ ]{%
        {\includegraphics[trim= 2 2 2 2, clip,width=0.25\textwidth, height=0.25\textwidth]{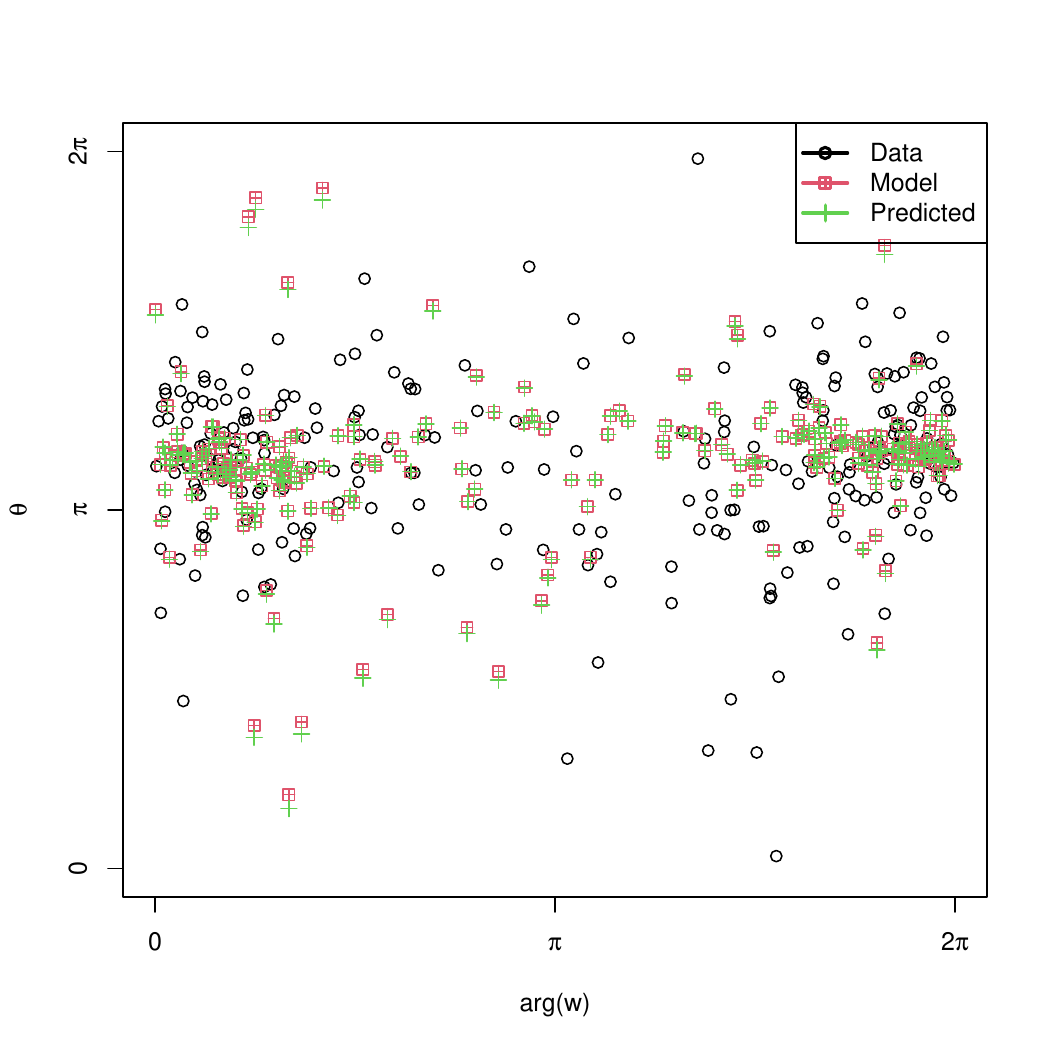}}}
   
    \subfloat[ ]{%
        {\includegraphics[trim= 2 2 2 2, clip,width=0.25\textwidth, height=0.25\textwidth]{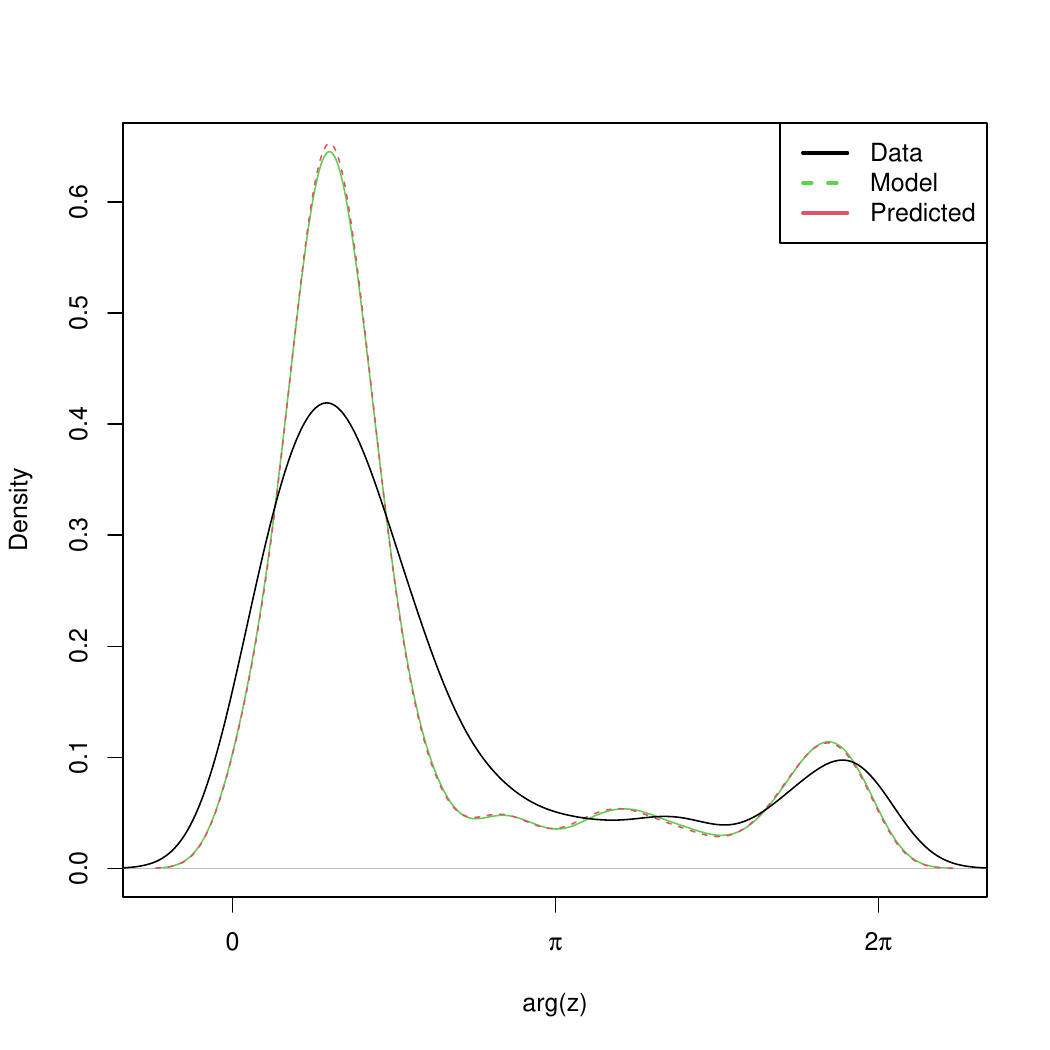}}}\vspace{5pt}
    \subfloat[ ]{%
        {\includegraphics[trim= 2 2 2 2, clip,width=0.25\textwidth, height=0.25\textwidth]{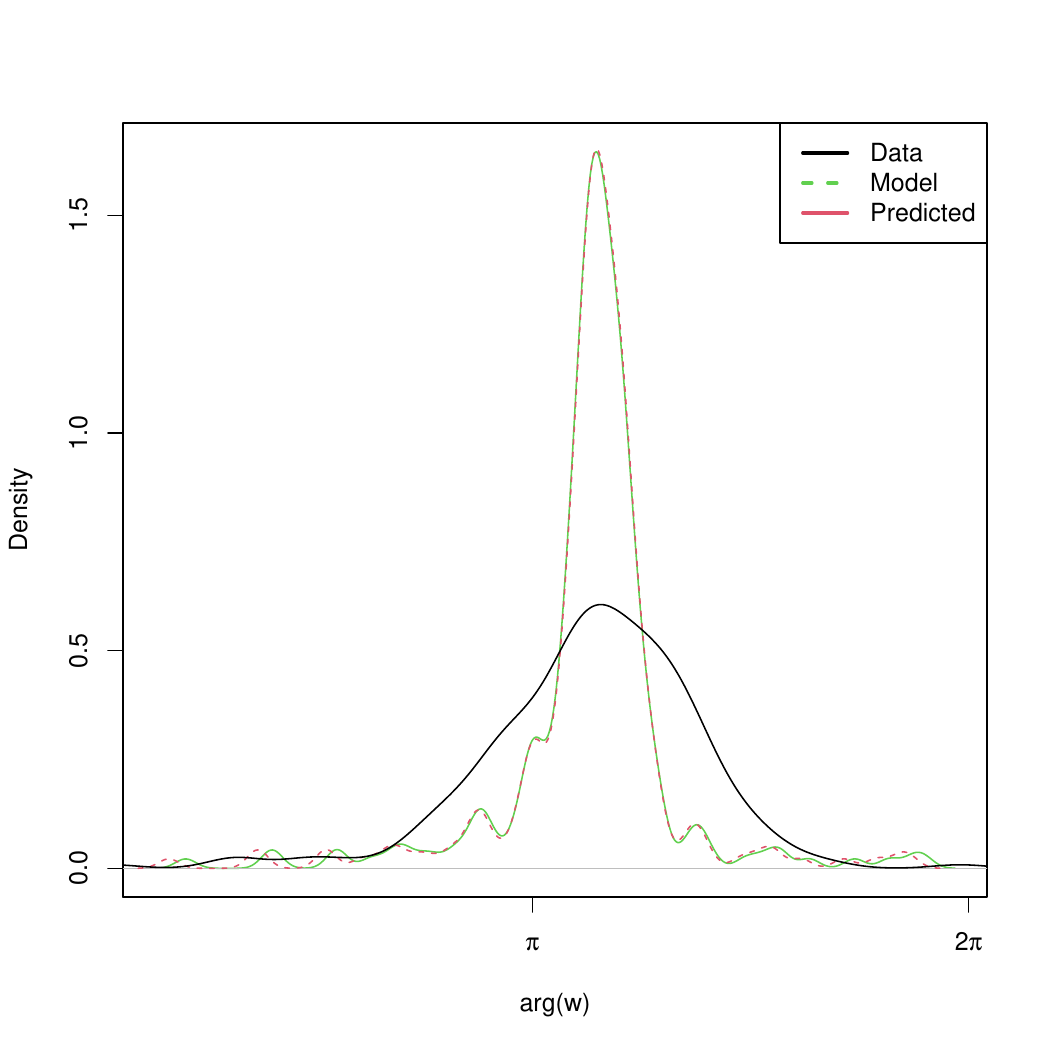}}}\vspace{5pt}
    \caption{ Scatter plot obtained from the model Equation-\ref{tor_reg_curve} (red box),  simulated data in Equation-\ref{tor_reg_model_full} (black small circle), and the predicted data (green cross) on the flat torus (a) for the horizontal angle $\phi$, and (b) for the vertical angle $\theta.$ 
(c) and (d) is the density plot of the same, respectively.}
     \label{simulated_plot_dens_scatter}
\end{figure*}

Figures \ref{simulated_plot_dens_scatter}(a) and (b) represent the scatter plot obtained from the model Equation-\ref{tor_mob_map},  simulated data in Equation-\ref{tor_reg_model_full}, and the predicted values for the horizontal angle $\phi$ and the vertical angle $\theta$, respectively.  In both figures, the x-axis denotes the arguments of the predictors $z$ and $w$, which are independently sampled from the von Mises distribution with the mean direction of 0 (in radians) and a concentration parameter $\kappa = 1$.  The y-axis denotes the response angles (in radians), derived from the arguments of the responses through the proposed torus-to-torus regression model.  Random angular errors are generated using a bivariate von Mises sine model with the zero mean direction, with marginal concentration parameters set at $\kappa_1 = \kappa_2 = 5$, and a dependency parameter of $\kappa_3 = 1$.
 Figures-\ref{simulated_plot_dens_scatter}(c) and (d) present the density plots for the horizontal angle $\phi$ and vertical angle $\theta$, respectively. 

 These plots illustrate the densities of the simulated data, the model with noise, and the predicted responses.  The plots collectively indicate that the proposed model effectively estimates parameters and recovers the underlying structure from the simulated dataset.

\begin{figure*}[t]
    \centering
    \subfloat[]{%
        {\includegraphics[trim= 2 2 2 2, clip,width=0.25\textwidth, height=0.25\textwidth]{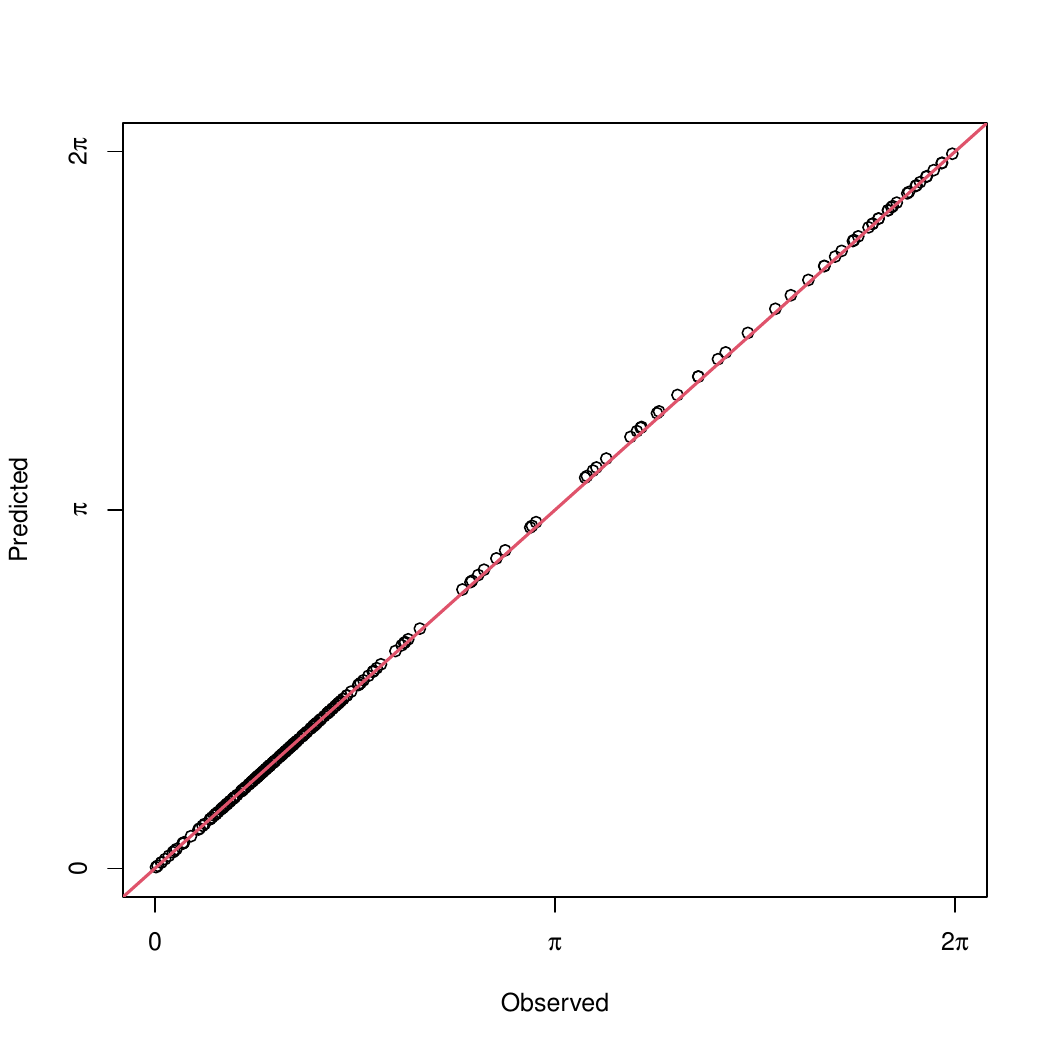}}}\hspace{5pt}
    \subfloat[ ]{%
        {\includegraphics[trim= 2 2 2 2, clip,width=0.25\textwidth, height=0.25\textwidth]{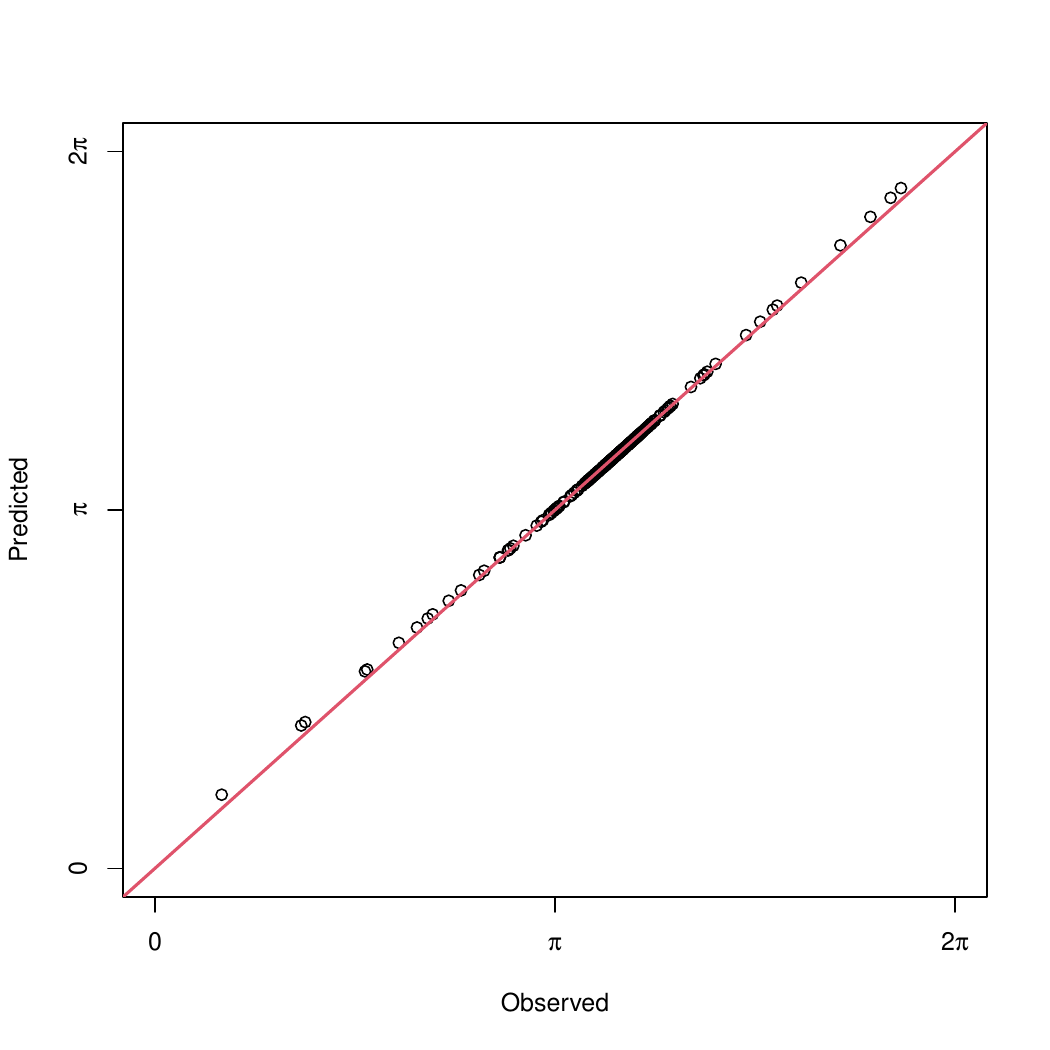}}}
   
    \subfloat[ ]{%
        {\includegraphics[trim= 2 2 2 2, clip,width=0.25\textwidth, height=0.25\textwidth]{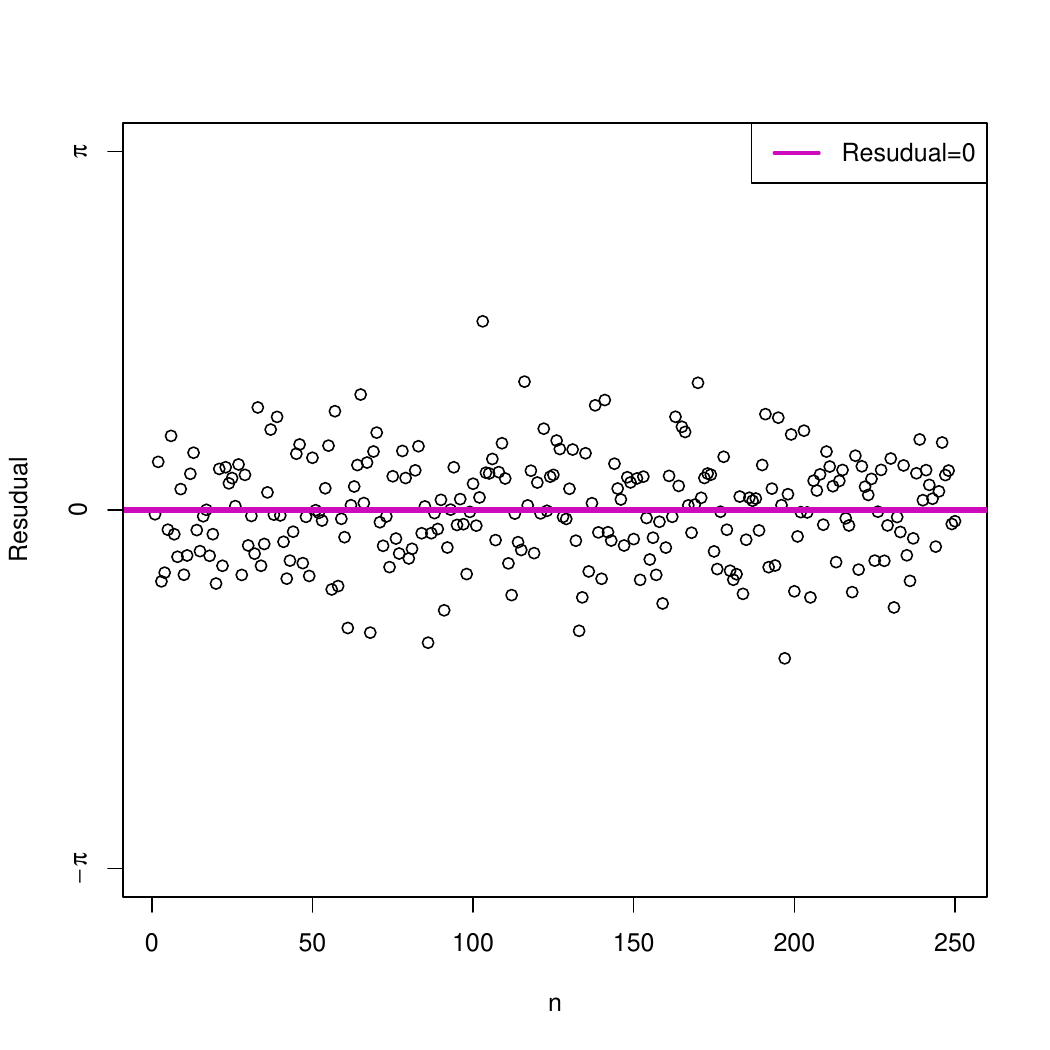}}}\vspace{5pt}
    \subfloat[ ]{%
        {\includegraphics[trim= 2 2 2 2, clip,width=0.25\textwidth, height=0.25\textwidth]{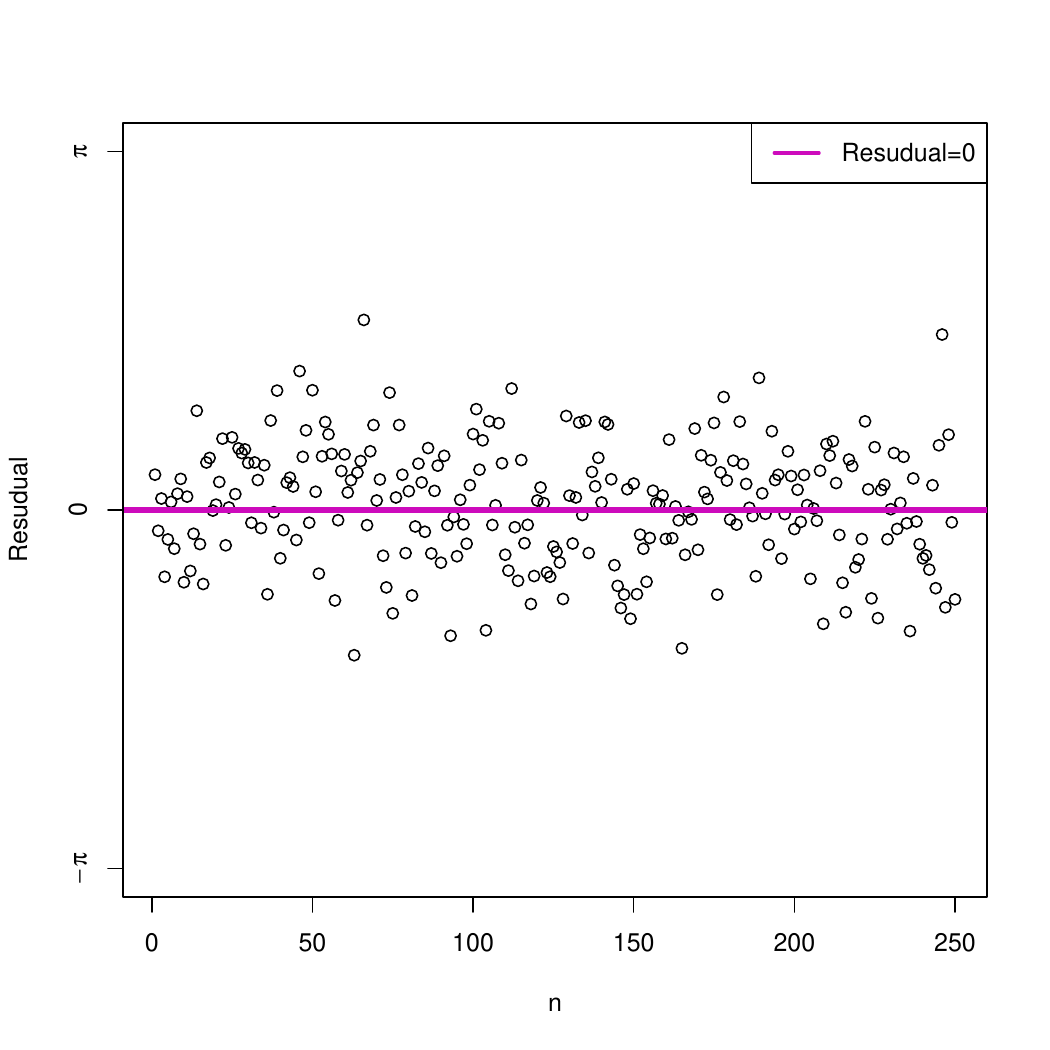}}}\vspace{5pt}
    \caption{ QQ plot of the observed and predicted (a) for the horizontal angle $\phi$, and (b) for the vertical angle $\theta.$ 
(c) and (d) is the plot of the residuals (restricted to the range \([-\pi, \pi]\) for enhanced visual clarity) with a reference line at  $0$ (radian).}
     \label{simulated_plot_qq_res}
\end{figure*}

Figures-\ref{simulated_plot_qq_res}(a) and (b) illustrate the quantile-quantile (QQ) plots for the horizontal angle $\phi$ and the vertical angle $\theta$, respectively.  The horizontal axis in these plots indicates the quantiles of the observed data, whereas the vertical axis illustrates the quantiles of the predicted values obtained from the proposed torus-to-torus regression model.  The proximity of the points to the $45^{\circ}$ reference line indicates that the model effectively represents the distributional properties of the simulated data.

Figures-\ref{simulated_plot_qq_res}(c) and (d) show the component-wise residual plots for $\phi$ and $\theta$, respectively.  The lack of a systematic pattern in these plots suggests that the angular errors are distributed randomly, supporting the effectiveness of the  model in representing the relationship between predictors and responses.  The residuals cluster near zero, aligning with the assumption of negligible angular error in a properly fitted regression model.
 Just as residuals in linear regression follow a normal distribution under well-specified models, the residuals in the proposed torus-to-torus regression framework are assumed to follow a von Mises (or circular normal) distribution  componentwise, specifically, for the horizontal angle $\phi$ and the vertical angle $\theta$.  This assumption was assessed using Watson’s test for circular normality or von Mises distribution.  The test statistic for the horizontal angle $\phi$ was 0.0318, which is below the 5\% critical value of 0.113.  Similarly, for the vertical angle $\theta$, the test statistic was 0.0325, which falls below the corresponding critical value of 0.117.  In both cases, the null hypothesis of von Mises-distributed residuals was not rejected, thus offering further support for the model's adequacy and its good fit to the simulated data.
   

 \subsection{Simulation Diagnostics for Branch-Cut Robustness}\label{sec:branch_cut}
In this section, we evaluate the numerical robustness of the torus‐to‐torus regression model with respect to the branch‐cut discontinuity in the complex power \(z^\eta\).  We perform three diagnostics: residual discontinuity and prediction continuity checks, cut‐placement sensitivity, and multiple random starts.
 (1) \textbf{Residual Discontinuity and Prediction Continuity:}
We compute circular residuals
$\delta_\phi \;=\;\bigl(\phi_{\rm pred}-\phi_{\rm obs}\bigr)\bmod2\pi,
~
\delta_\theta \;=\;\bigl(\theta_{\rm pred}-\theta_{\rm obs}\bigr)\bmod2\pi,$
and plot them against the predictor angles \(\arg(z)\) and \(\arg(w)\).  As shown in Figure-\ref{simulated_plot_qq_res}(c) and (d), neither residual exhibits large spikes near \(\pi\) or \(-\pi\), and the scatter of \(\delta_\phi\) and \(\delta_\theta\) remains uniform across the full \(2\pi\) range.  The QQ‐style plots of predicted versus observed angles lie  on a line passing through the origin with a slope $\pi/4$ with no jumps. This confirms smooth prediction continuity.
(2) \textbf{Cut-Placement Sensitivity:}\label{sec:sim-cut}
To test invariance to the location of the branch‐cut, we rotated both circular predictors by angles \(t\in\{0,\pi/6,\dots,11\pi/6\}\) and refitted the model. The true parameters for the model in Equation-\ref{tor_reg_model} in this case are $(\phi_0, b_1, b_2, b_3, b_4, \theta_0)=(\pi/3, 1.3, 0.5, -0.7, 0.4,\pi, 0.5)$. The angular errors derived from a von Mises sine model with $\kappa_1=\kappa_2=4,$ and $\kappa_3=1$, and the 250 covariates are drawn from a von Mises sine model as well with $\kappa_1=\kappa_2=2,$ and $\kappa_3=-1$.
Table-\ref{tab:cut_placement_sensitivity} reports the maximum relative change in each parameter over all rotations.  All the parameter changes are well below the 5\% threshold.  This demonstrates that parameter estimates are effectively invariant to the arbitrary choice of branch cut.
(3) \textbf{Multiple Random Starts:}\label{sec:sim-starts}
We ran the optimizer from multiple random initializations, each generated by adding independent Gaussian noise to the true parameter vector in the simulation. 
Together, these diagnostics, supported by Figure-\ref{simulated_plot_qq_res}, Table-\ref{tab:cut_placement_sensitivity}, and the multiple-start analysis, establish that despite the theoretical branch‐cut in \(z^\eta\), the regression model, it remains numerically robust and statistically reliable across the full torus.

\begin{table*}[h!]
\centering
\resizebox{0.7\textwidth}{!}{%
\begin{tabular}{c|ccccccc}
\toprule
Rotation (°) & $\Delta\phi_0$ (\%) & $\Delta\Re(\beta_1)$ (\%) & $\Delta\Im(\beta_1)$ (\%) & $\Delta\Re(\beta_2)$ (\%) & $\Delta\Im(\beta_2)$ (\%) & $\Delta\theta_0$ (\%) & $\Delta\eta$ (\%) \\
\midrule
30  & 0.79 & 0.08 & 0.40 & 0.14 & 0.43 & 0.01 & 0.46 \\
60  & 1.77 & 0.00 & 1.00 & 0.20 & 0.63 & 0.13 & 0.56 \\
90  & 1.63 & 0.37 & 0.94 & 0.01 & 0.15 & 0.15 & 0.44 \\
120 & 1.30 & 0.22 & 0.70 & 0.71 & 0.18 & 0.02 & 0.59 \\
150 & 0.28 & 0.15 & 0.05 & 0.47 & 0.17 & 0.75 & 0.64 \\
180 & 0.35 & 0.17 & 0.22 & 0.54 & 0.10 & 0.68 & 0.65 \\
210 & 0.13 & 0.77 & 0.91 & 0.45 & 0.03 & 0.49 & 0.82 \\
240 & 0.29 & 0.33 & 0.59 & 0.36 & 0.18 & 0.82 & 0.88 \\
270 & 1.44 & 0.52 & 1.28 & 0.49 & 0.42 & 0.94 & 0.70 \\
300 & 1.10 & 1.38 & 1.07 & 0.57 & 0.29 & 0.78 & 0.56 \\
330 & 0.76 & 1.23 & 0.47 & 0.56 & 0.48 & 0.38 & 0.47 \\
\midrule
\textbf{Max} & \textbf{1.77} & \textbf{1.38} & \textbf{1.28} & \textbf{0.71} & \textbf{0.63} & \textbf{0.94} & \textbf{0.88} \\
\bottomrule
\end{tabular}%
}
\vspace{0.3cm}
\caption{Cut-placement sensitivity: relative percentage ranges compared to baseline (No Rotation). For no rotation, the estimated values of the parameters are $(\hat{\phi}_0, \hat{b}_1, \hat{b}_2, \hat{b}_3, \hat{b}_4, \hat{\theta}_0, \hat{\eta}) =(1.0465, 1.3008, 0.5080, -0.7036, 0.3975, 3.1158, 0.4947 )$.}
\label{tab:cut_placement_sensitivity}
\end{table*}

\subsection{ Comparison}
 To the best of our knowledge, there is no torus-to-torus regression model in the literature. The proposed torus-to-torus regression model (Equation-\ref{tor_mob_map}) generalizes existing circular-circular regression models that use a single M\"{o}bius transformation as the link function. In particular,  the models of \cite{jha2017multiple, jha2018circular} \cite{downs2002circular}, and \cite{kato2008circular} all regress a single circular response on a single circular predictor via the M\"{o}bius map (Equation-\ref{cir_mob_map}). The proposed model extends this framework to bivariate angular predictors and bivariate angular responses by coupling two M\"{o}bius components through the interaction parameter $\eta$, which captures cross-predictor nonlinear dependence on the torus.
A further distinction lies in the estimation philosophy.  All the above-mentioned models are fully parametric, assuming wrapped Cauchy or von Mises errors. In contrast, the proposed model is semi-parametric; it makes no distributional assumption on the angular error and instead minimizes a geometric loss (Equation-\ref{loss function total}) derived from the intrinsic geometry of the torus and the sphere.

\subsection{Circular Scatter Plots} \label{csctr_plot}
Traditional Cartesian scatter plots of angular data (e.g., plotting observed vs. predicted angles in an $x$-$y$ coordinate system) are not appropriate in circular settings due to their sensitivity to the arbitrary choice of the zero angle. Alternatives such as the spoke-plot and donut-plot, introduced by \cite{zubairi2008alternative} and \cite{jha2017multiple}, offer improved visualization for circular data by aligning with its geometric nature. However, these methods also have limitations. For instance, spoke plots can become cluttered and difficult to interpret when the sample size is large, as numerous intersecting lines obscure the underlying structure. Similarly, donut plots, which use two concentric circles to depict observed and predicted values, may lead to confusion due to visual complexity, especially when overlaid with dense data points.

To address these limitations, we propose a new visualization method termed the \textit{Circular Scatter Plot} for representing angular data. Suppose we have $n$ angular observations $\theta_1, \ldots, \theta_n$. We construct $n$ concentric circles centered at the origin, with the radius of the $i$-th circle set to $\frac{i}{n}$, such that the innermost circle corresponds to  index $1$ and the outermost to index $n$. Each observation $\theta_i$ is then plotted on its corresponding circle at the Cartesian coordinates $\frac{i}{n}(\cos\theta_i, \sin\theta_i)$.

One of the key advantages of the proposed Circular Scatter Plot is its ability to preserve the exact angular location of each data point, analogous to how a Cartesian scatter plot in Euclidean space preserves precise spatial information. Additionally, the proximity between observed and predicted points provides a direct visual cue for assessing model fit: closely clustered observed and predicted points indicate a good fit, whereas significant separation suggests model inadequacy or poor parameter estimation.

\section{Data Analysis}\label{data_analysis}
 The super-cyclones  \textit{Biparjoy} (June 2023) and \textit{Amphan} (May 2020) underscore the critical need to understand and address the interplay between wind direction and wave direction during extreme weather events. In both cases, the cyclones exhibited intense and sustained wind fields.  \textit{Biparjoy} reached peak wind speeds of 195 km/h, and \textit{Amphan} sustained winds up to 165 km/h with gusts of 185 km/h at landfall. These high-velocity winds exerted tremendous force on the ocean surface, transferring energy that generated large, coherent wave fields. However, the directionality of these winds relative to the coastline played a pivotal role in determining the severity and nature of coastal impacts. For instance, during  \textit{Biparjoy}, the alignment of wind and wave directions contributed significantly to severe coastal flooding and erosion along Gujarat’s coast. Similarly, in \textit{Amphan}, as the storm approached the West Bengal coast, the strong onshore wind components drove waves directly toward the coastline, exacerbating storm surge and inland inundation. 
Intending to study the possible association of the wind direction and wave direction, we collected the 10-meter above the sea level wind direction and mean-wave direction data \cite[see][]{data2023}.

For   Cyclone  \textit{Biparjoy}, the data were collected from a fixed location at coordinates $17.3^{\circ} N$, $67.3^{\circ} E$ (near eye), situated in the east-central Arabian Sea. Measurements of both wind direction and mean wave direction were recorded twice daily at 6:00 AM and 12:00 PM (noon) from   1st May to 30th July 2023. This two-month period spans not only the active duration of the cyclone (6th-19th June 2023) but also includes several weeks before and after the event, providing a broader context for comparative analysis.
Similarly, for   Cyclone \textit{Amphan}, data were gathered from the near landfall location, $21.5 ^\circ N, 88.5^\circ E$ in the southeastern Bay of Bengal. Wind and wave direction readings were recorded at 7:00 AM and 1:00 PM daily, spanning the period from   1st April to 30th June 2020. This time frame encompasses the lifecycle of \textit{Amphan} (16th-21st May 2020) as well as days before and after the cyclone. The inclusion of both pre-cyclonic and post-cyclonic periods enables a comparative study of directional coherence under both normal and extreme weather conditions.

\begin{figure*}[h!]
    \centering
    \subfloat[]{%
        {\includegraphics[trim= 2 2 2 2, clip,  width=0.33\textwidth, height=0.33\textwidth]{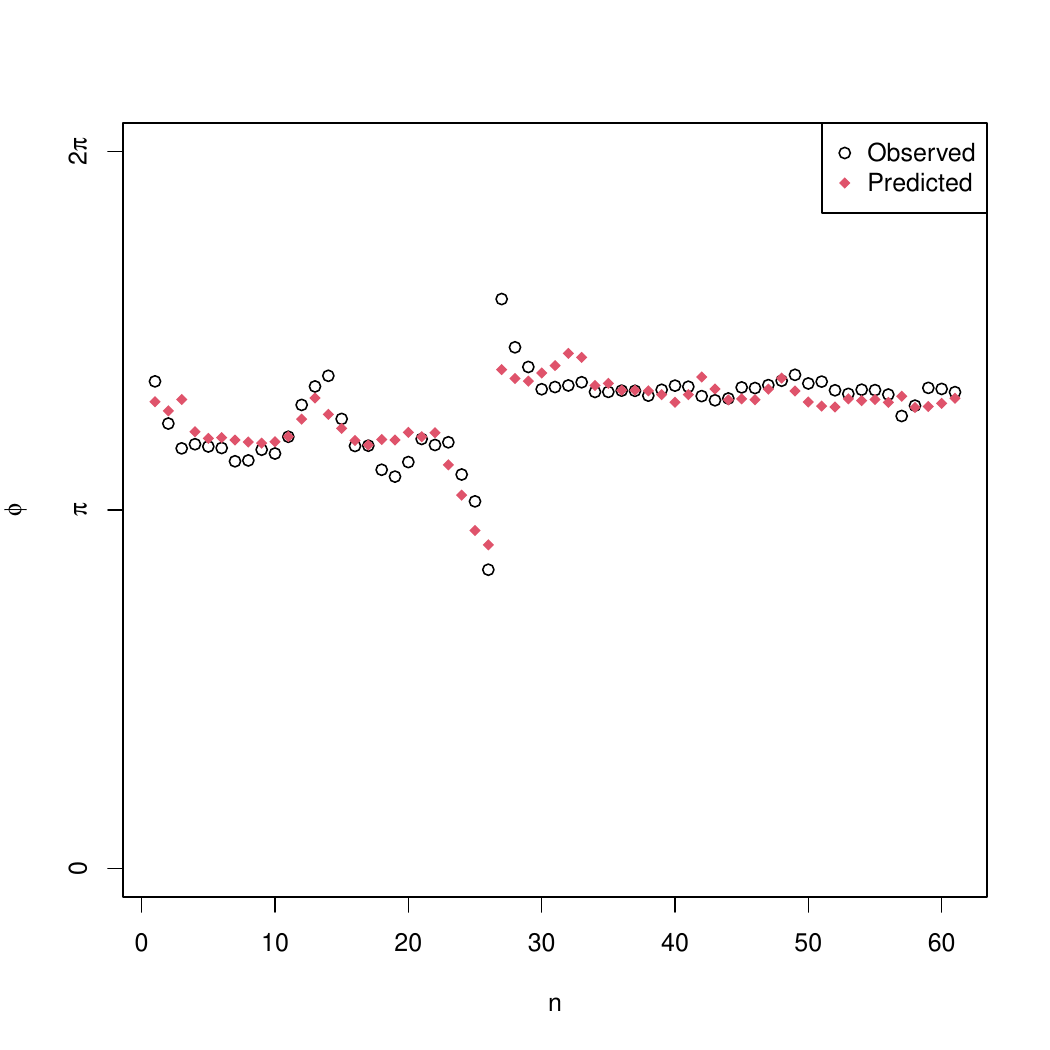}}}
        \subfloat[ ]{%
        {\includegraphics[trim= 2 2 2 2, clip,  width=0.33\textwidth, height=0.33\textwidth]{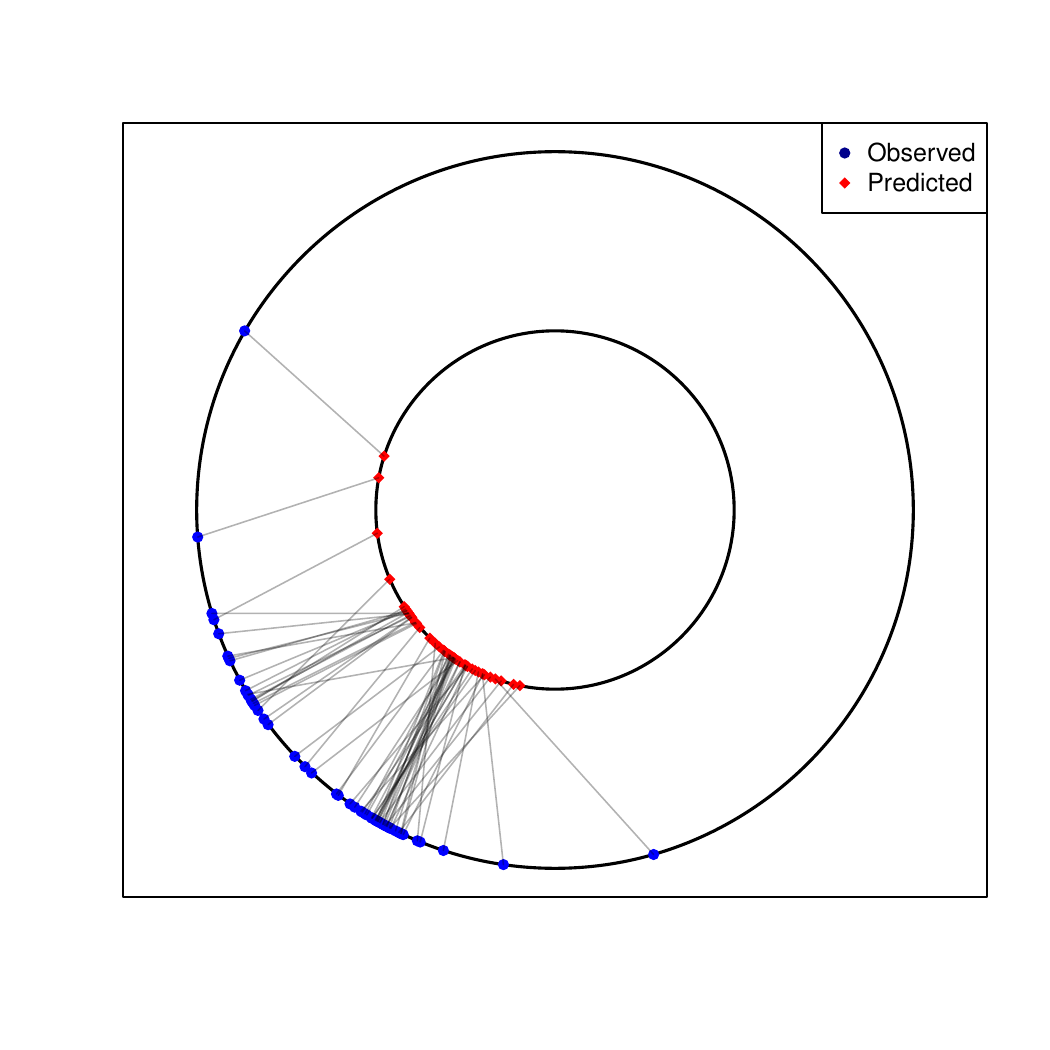}}}
        \subfloat[ ]{%
        {\includegraphics[trim= 2 2 2 2, clip,  width=0.33\textwidth, height=0.33\textwidth]{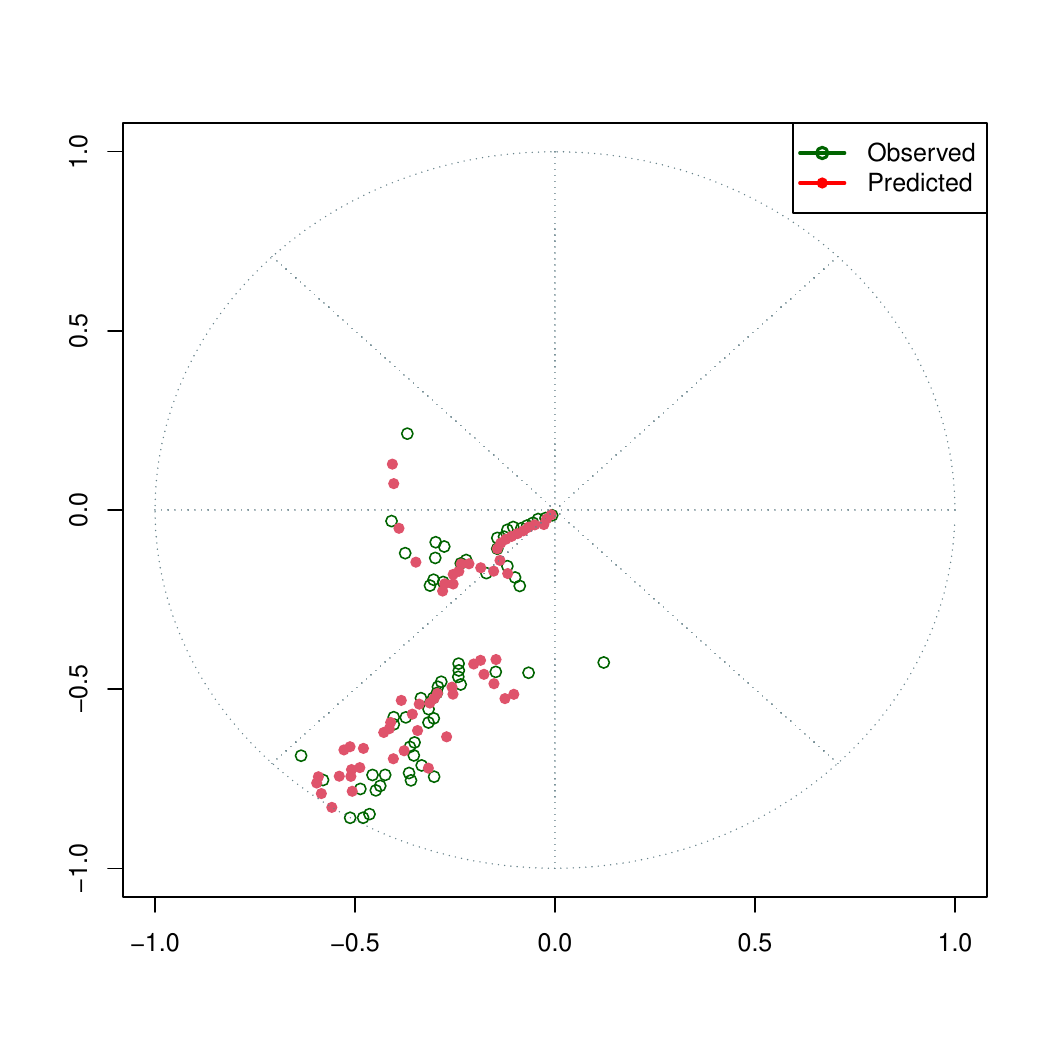}}}
        
     \subfloat[ ]{%
        {\includegraphics[trim= 2 2 2 2, clip,  width=0.33\textwidth, height=0.33\textwidth]{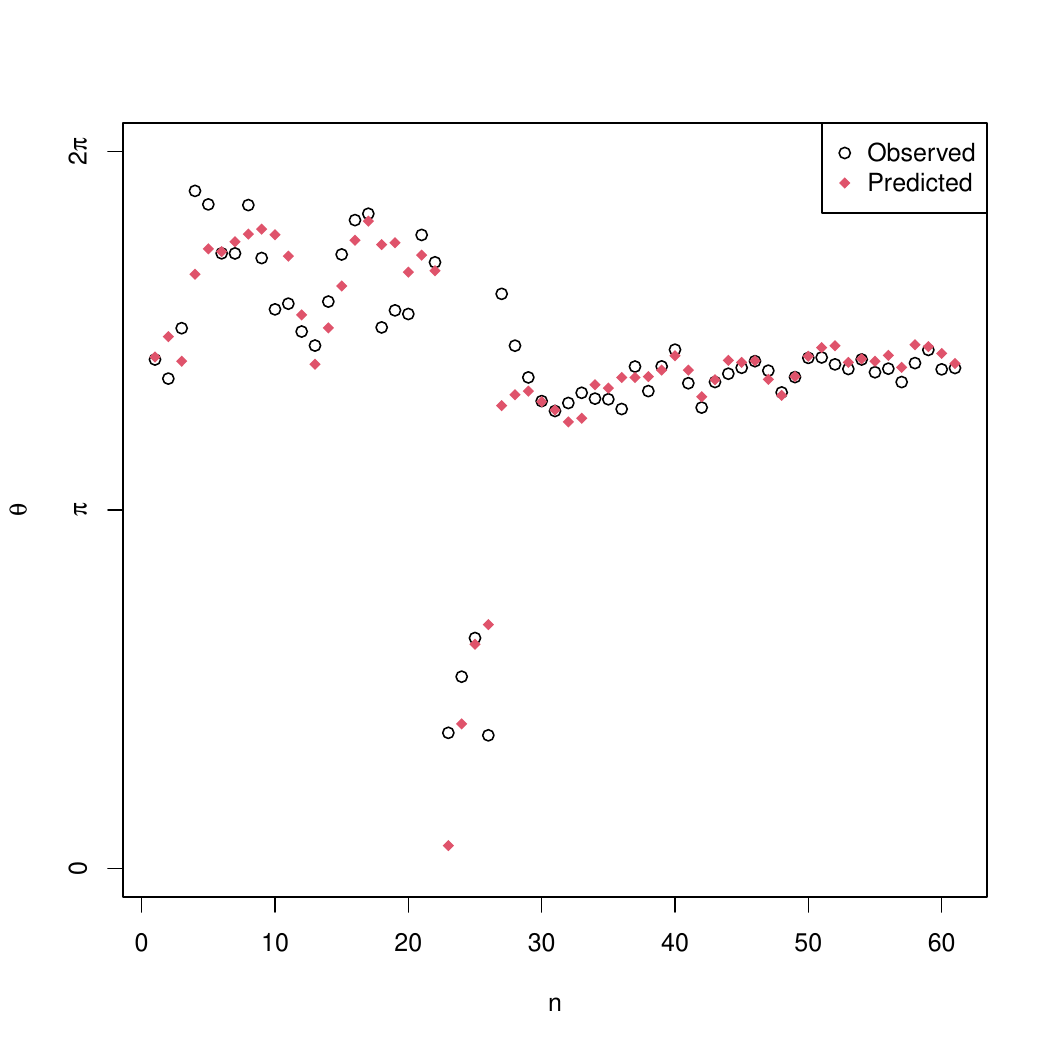}}}
         \subfloat[ ]{%
        {\includegraphics[trim= 2 2 2 2, clip,  width=0.33\textwidth, height=0.33\textwidth]{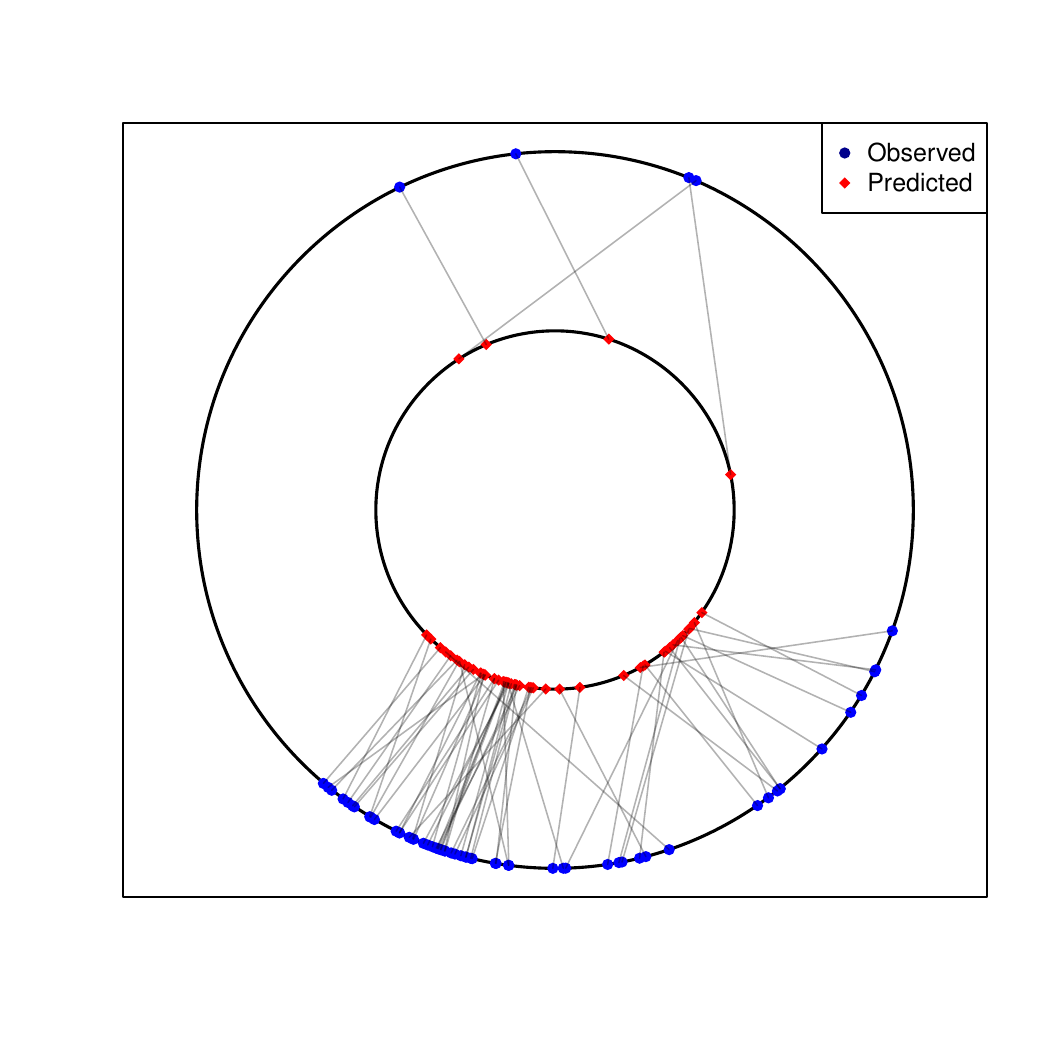}}}
    \subfloat[ ]{%
        {\includegraphics[trim= 2 2 2 2, clip,  width=0.33\textwidth, height=0.33\textwidth]{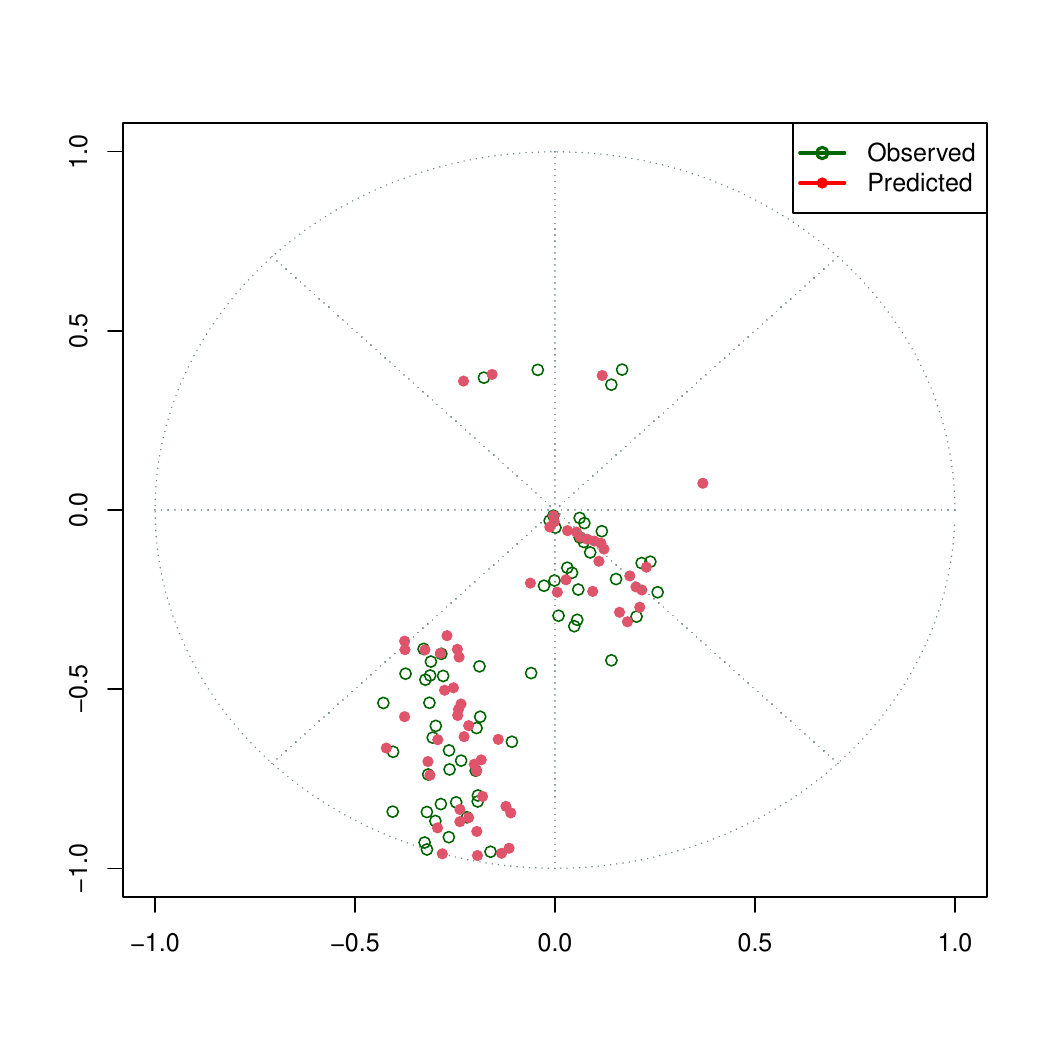}}}
        \hspace{3pt}
 \caption{For  \textit{Biparjoy}: (a) and (d) display the scatter plots of the observed and predicted wind and wave directions, respectively. Figures (b) and (e) present the corresponding spoke plots, where the outer circle represents the observed directions and the inner circle represents the predicted directions. Finally, Figures (c) and (f) illustrate the newly proposed circular scatter plots for the observed and predicted wind and wave directions, respectively. These visualizations help assess the accuracy and quality of the model predictions.}
     \label{data_analysis_bip}
\end{figure*}

 For both events, wind and wave directions recorded in the morning (6:00 AM for  \textit{Biparjoy}, 7:00 AM for \textit{Amphan}) were treated as covariates, while those recorded in the afternoon (12:00 PM for  \textit{Biparjoy}, 1:00 PM for \textit{Amphan}) served as responses. Here,  wave direction is represented as $\phi$, and wind direction is represented as $\theta.$ The parameter estimates of the proposed regression model, including standard errors, are summarized in Table-\ref{table:data_bip_amp_table}.
To estimate the model parameters, we employed a numerical optimization method to minimize the loss function defined in Equation-\ref{loss function total}.
Given the complex nature of the loss function and the risk of the optimization process converging to local minima, we ran the optimization procedure 1000 times, each with different randomly initialized starting values. Specifically, the parameters $\phi_0$ and $\theta_0$ were initialized from a Uniform distribution over $[0, 2\pi]$, and the parameters $b_1, b_2, b_3,$ and $b_4$ were initialized from a Uniform distribution over $[-5, 5]$.
From these runs, we selected only those parameter estimates that resulted in relatively low standard errors and yielded satisfactory model diagnostics, particularly in terms of visual assessment through the 
spoke-plot and the proposed circular scatter plots in Section-\ref{csctr_plot}.

\begin{figure*}[h!]
    \centering
    \subfloat[]{%
        {\includegraphics[trim= 2 2 2 2, clip,  width=0.33\textwidth, height=0.33\textwidth]{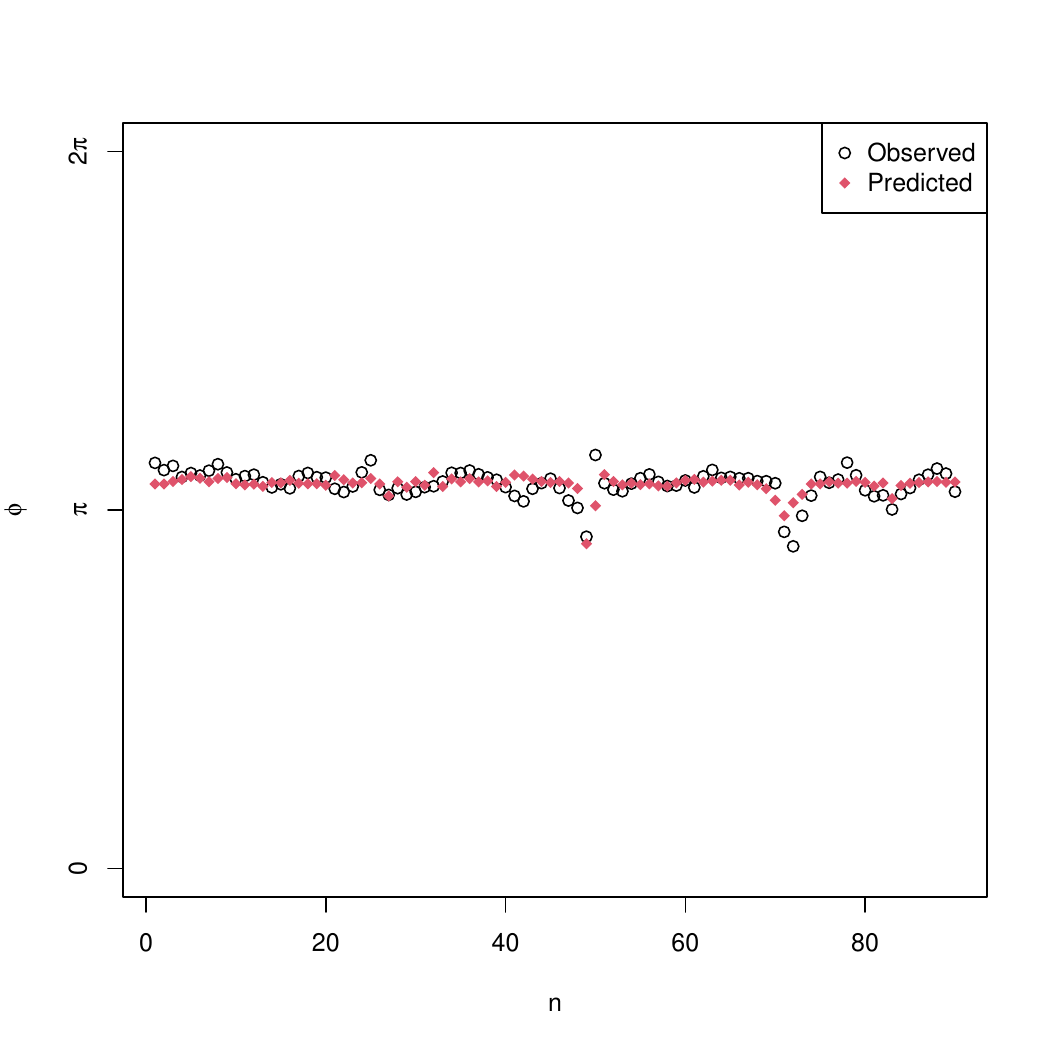}}}
    \subfloat[ ]{%
        {\includegraphics[trim= 2 2 2 2, clip,  width=0.33\textwidth, height=0.33\textwidth]{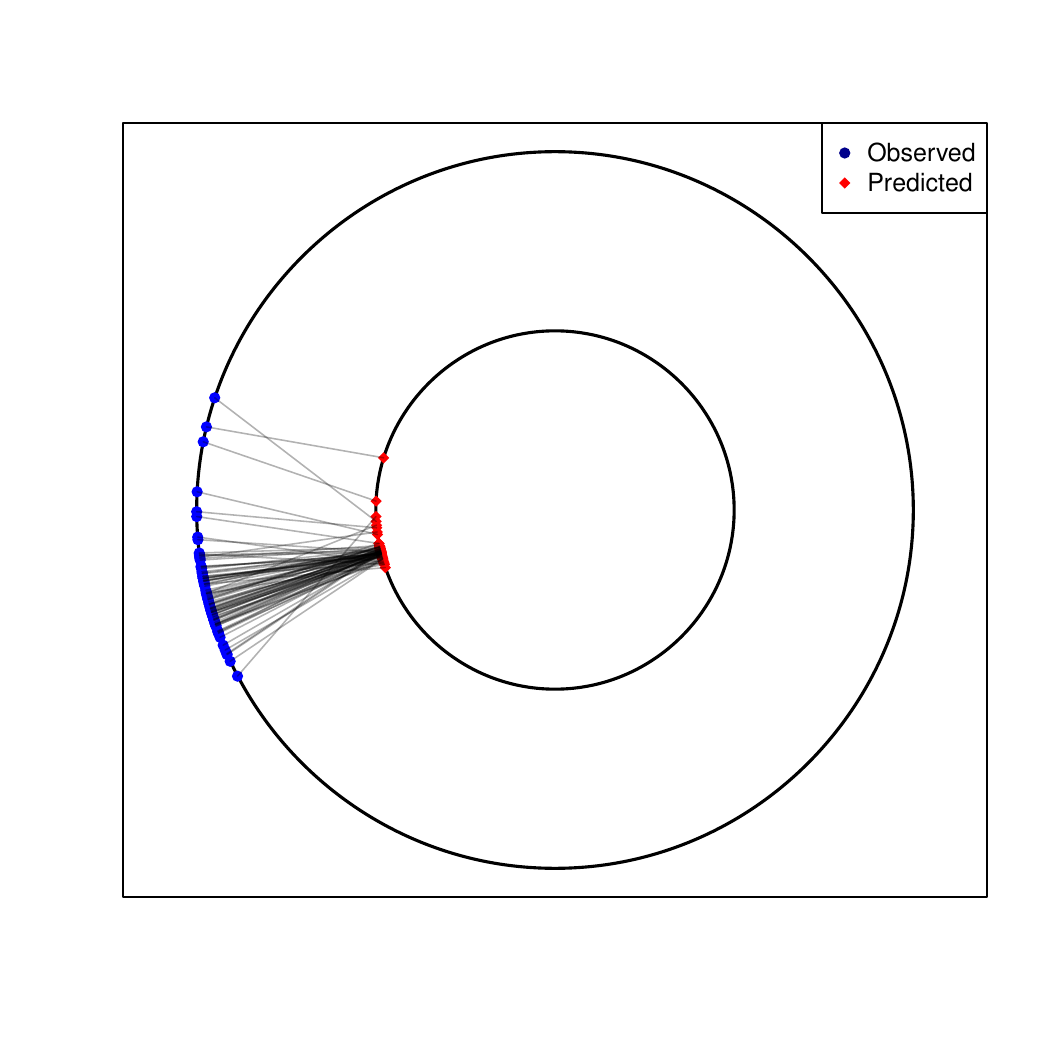}}}
    \subfloat[ ]{%
        {\includegraphics[trim= 2 2 2 2, clip,  width=0.33\textwidth, height=0.33\textwidth]{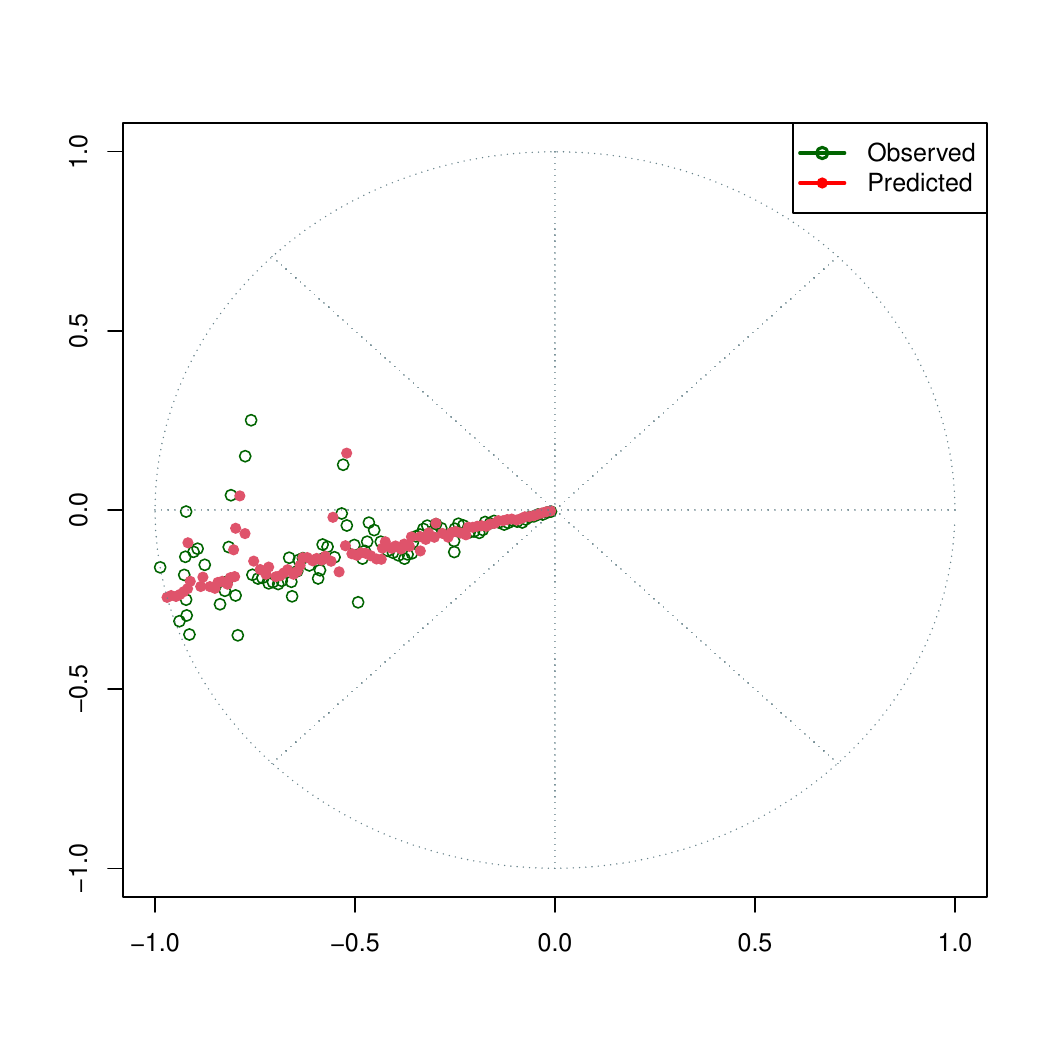}}}
        
    \subfloat[ ]{%
        {\includegraphics[trim= 2 2 2 2, clip,  width=0.33\textwidth, height=0.33\textwidth]{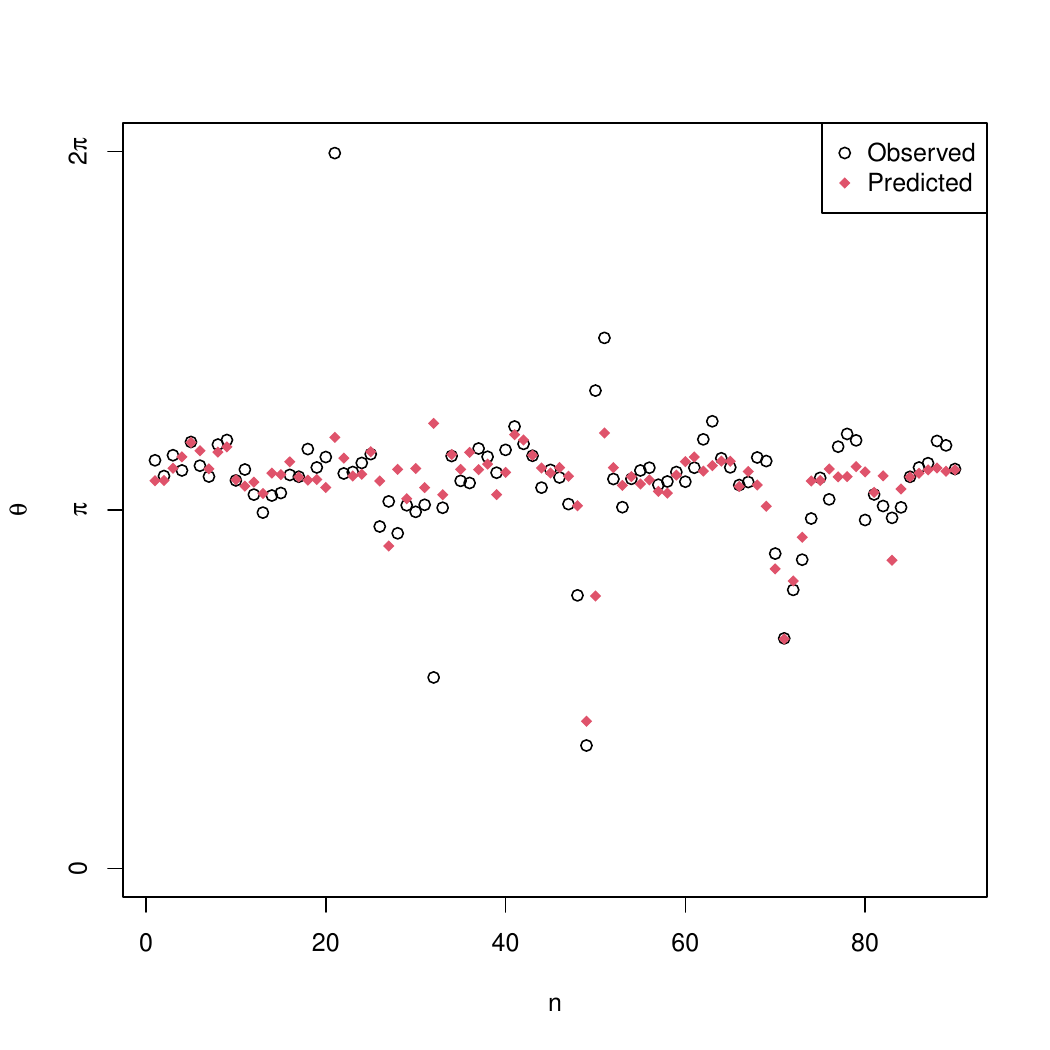}}}
          \subfloat[ ]{%
        {\includegraphics[trim= 2 2 2 2, clip,  width=0.33\textwidth, height=0.33\textwidth]{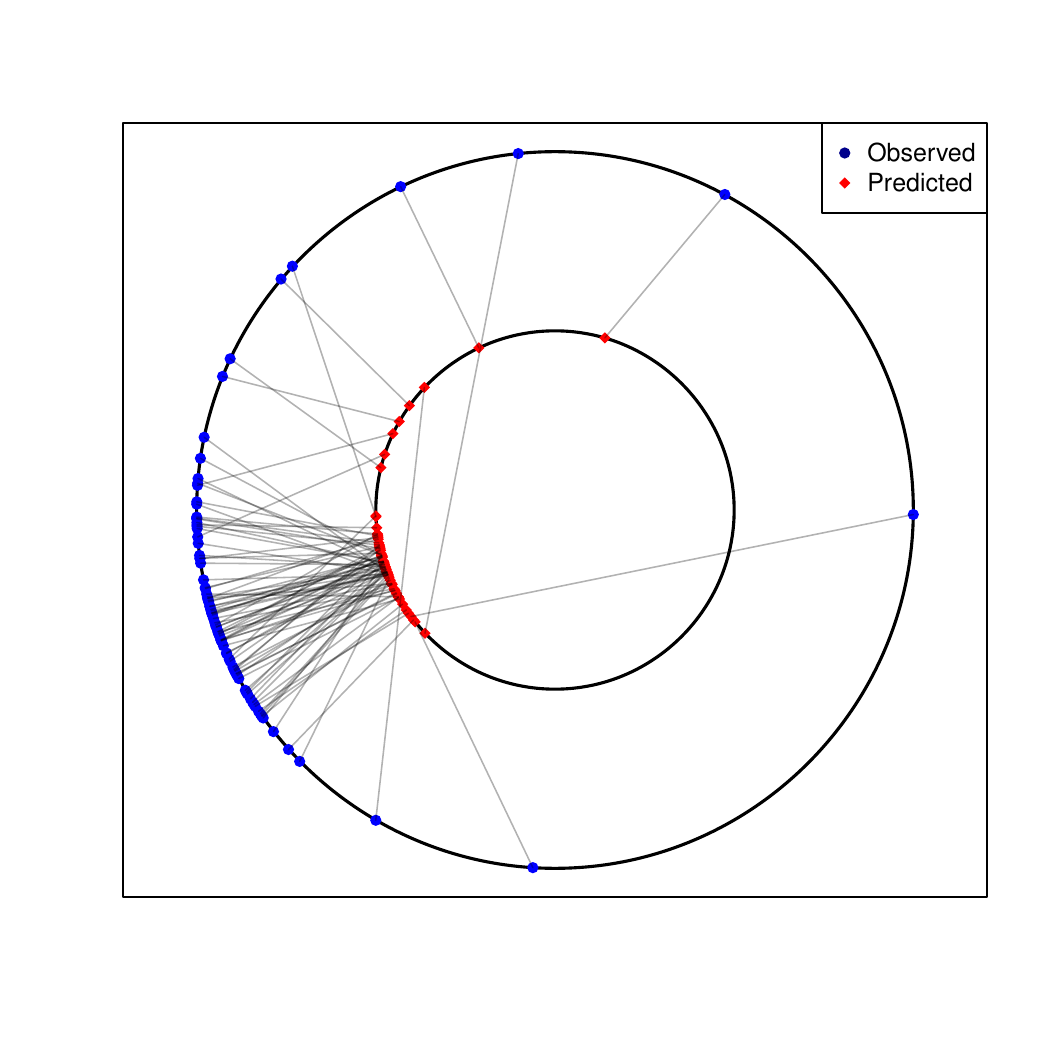}}}
    \subfloat[ ]{%
        {\includegraphics[trim= 2 2 2 2, clip,  width=0.33\textwidth, height=0.33\textwidth]{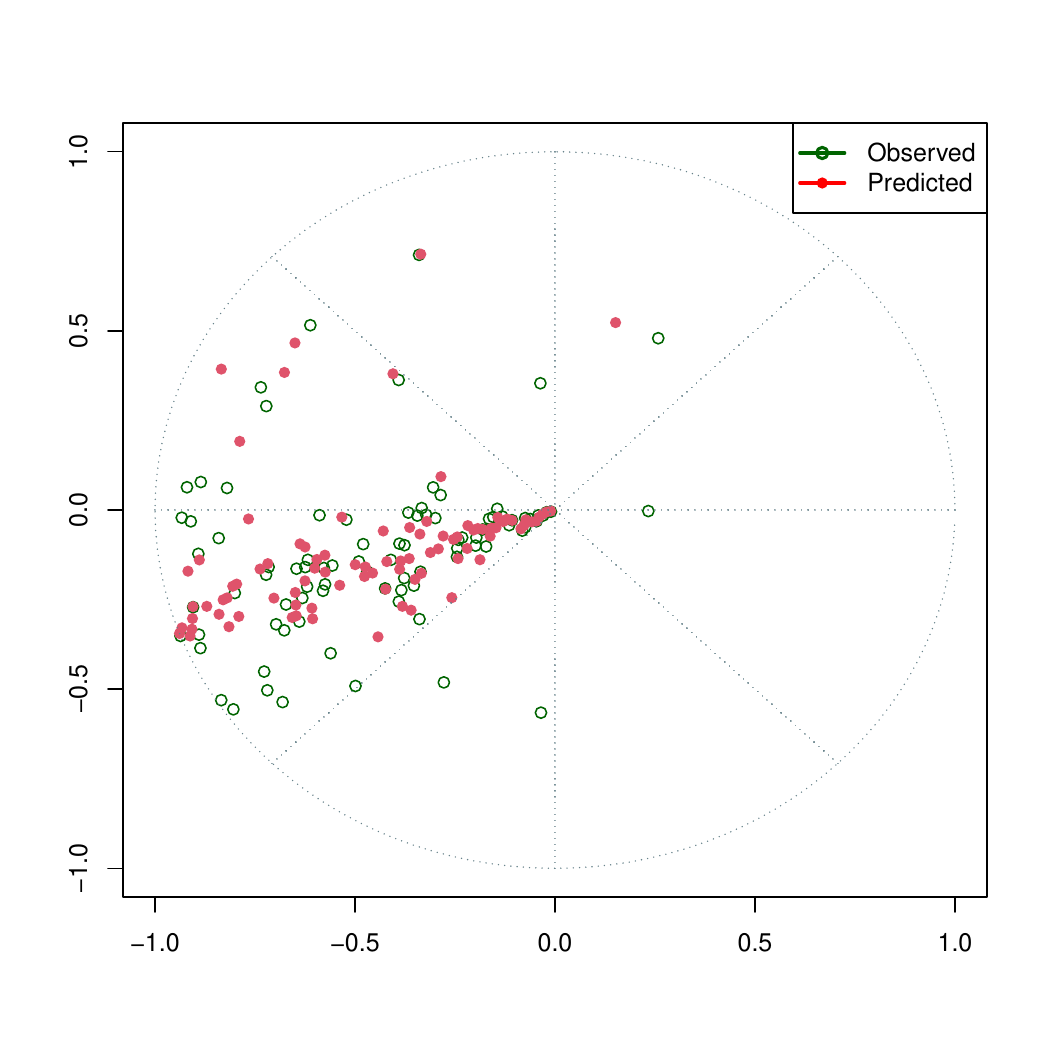}}}
    \caption{For \textit{Amphan}: Panels (a) and (d) show scatter plots for wind and wave directions, respectively. Panels (b) and (e) present spoke-plots, where the outer ring denotes the observed directions and the inner ring represents the predicted ones. Panels (c) and (f) feature the circular scatter plots developed in this study.}
     \label{data_analysis_amp}
\end{figure*}

\begin{table*}[b]
\centering
\renewcommand{\arraystretch}{1} 
\begin{tabular}{ |>{\centering\arraybackslash}p{1.6cm}|
                  >{\centering\arraybackslash}p{1.6cm}|
                  >{\centering\arraybackslash}p{1.6cm}|
                  >{\centering\arraybackslash}p{1.6cm}|
                  >{\centering\arraybackslash}p{1.6cm}|
                  >{\centering\arraybackslash}p{1.6cm}|
                  >{\centering\arraybackslash}p{1.6cm}|
                  >{\centering\arraybackslash}p{1.6cm}| }
  \hline
Super Cyclone (location) & 
$\hat{\phi}_0~(s.e)$ in radian & 
$\hat{b}_1~(s.e)$ & 
$\hat{b}_2~(s.e)$ & 
$\hat{b}_3~(s.e)$ & 
$\hat{b}_4~(s.e)$ & 
$\hat{\theta}_0~(s.e)$ in radian & 
$\hat{\eta}~(s.e)$ \\ 
  \hline\hline  
 \textit{Biparjoy} (eye) &  
6.2604 (0.0447) & 
-0.6852 (0.4163) &  
-0.3576 (0.0481) & 
-0.0672 (0.0045) & 
0.1507 (0.0180)& 
3.6616 (0.0174) & 
1.1388 (0.0220) \\ 
  \hline

 
\textit{Amphan} (landfall)& 
 4.1082 (0.0351) & 
0.9017 (0.3750) &  
-0.7849 (0.0142) & 
0.2165 (0.0201) & 
0.2923 (0.0354)& 
2.6826 (0.0415) & 
1.1391 (0.0242) \\
  \hline
\end{tabular}
\vspace{0.3cm}
\caption{Estimated parameter values (with standard errors) for  \textit{Biparjoy} and \textit{Amphan} cyclone data.}
\label{table:data_bip_amp_table}
 \end{table*}

        

To evaluate the performance of the proposed model, we employed several visual diagnostic tools. For    \textit{Biparjoy}, scatter plots in Figures-\ref{data_analysis_bip}(a) and (d) compare the observed and predicted wind and wave directions, respectively, highlighting the pointwise agreement between the two. The spoke plots in Figures-\ref{data_analysis_bip}(b) and (e) offer a circular representation where the outer ring denotes observed values and the inner ring represents model predictions, enabling an intuitive comparison of angular alignment. Additionally, Figures-\ref{data_analysis_bip}(c) and (f) showcase the newly introduced   Circular Scatter Plot, which maps each observation onto a series of concentric circles. This visualization allows for a clear assessment of both temporal progression and the directional fit of the model.
A parallel analysis was conducted for   \textit{Amphan}. Scatter plots in Figures-\ref{data_analysis_amp}(a) and (d) again illustrate the agreement between observed and predicted directions for wind and wave data. The spoke plots in Figures-\ref{data_analysis_amp}(b) and (e) similarly provide a radial comparison of the two sets of values. Notably, the circular scatter plots can be seen in Figures-\ref{data_analysis_amp}(c) and (f).

The application of the proposed torus-to-torus regression model to wind and wave direction data from the super cyclones  \textit{Biparjoy} and \textit{Amphan} demonstrated its effectiveness in capturing complex directional relationships in meteorological phenomena. For Cyclone  \textit{Biparjoy}, the model identified a significant interaction between morning and afternoon wind-wave directions, with an estimated interaction parameter $\hat{\eta} = 1.1390 $, indicating a meaningful joint influence of the bivariate angular predictors on the response. Similarly, the analysis of Cyclone \textit{Amphan} yielded an estimate of $\hat{\eta} = 1.1391 $, suggesting it also has an interaction effect between the angular variables during this event. Visual diagnostic tools, including scatter plots, spoke plots, and circular scatter plots, confirmed a close agreement between observed and predicted values for both cyclones. These findings demonstrate the model’s flexibility and adaptability to differing meteorological conditions, providing valuable insights into the directional dynamics of wind and wave interactions during extreme cyclonic events along India’s eastern and western coasts. Across both events, the proximity of observed and predicted points in these visualizations highlights the model’s strong capability to effectively capture underlying directional patterns. The results suggest that the torus-to-torus regression framework is especially well-suited for modeling paired circular data in complex meteorological settings, where wind and wave directions are interrelated and evolve nonlinearly over time. Ultimately, this approach affirms the importance of flexible directional regression models and underscores their potential to enhance cyclone analysis, forecasting, and risk mitigation in coastal and marine environments.

\subsection{Geophysical Interpretation of Estimated Parameters}
The estimated Möbius parameters from Table-\ref{table:data_bip_amp_table} reveal distinct wind-wave directional patterns for  \textit{Biparjoy} and \textit{Amphan} cyclones.\\
\textbf{ \textit{Biparjoy} (Arabian Sea):} 
$\beta_1 \approx -0.6852 - 0.3576i$ yields $|\beta_1| = 0.77$ and $\arg(\beta_1) = 208^\circ$, while $\gamma_1 \approx -0.0672 + 0.1507i$ gives $|\gamma_1| = 0.17$ and $\arg(\gamma_1) = 114^\circ$. The inequality $|\beta_1| > |\gamma_1|$ and $94^\circ$ argument separation indicate a stronger wave directional response relative to winds.

\noindent
\textbf{\textit{Amphan} (Bay of Bengal):} 
$\beta_1 \approx 0.9017 - 0.7849i$ yields $|\beta_1| = 1.20$ and $\arg(\beta_1) = 319^\circ$, while $\gamma_1 \approx 0.2165 + 0.2923i$ gives $|\gamma_1| = 0.36$ and $\arg(\gamma_1) = 53^\circ$. Here $|\beta_1| > |\gamma_1|$ persists with $-95^\circ$ circular separation (equivalent to $265^\circ$), which again indicates a stronger wave directional response relative to winds.

Across both cyclones, $|\beta_1| > |\gamma_1|$ consistently holds. The distinct argument separations ($94^\circ$ vs. $-95^\circ$) and quadrant placements (IV-II for  \textit{Biparjoy}; IV-I for \textit{Amphan}) capture basin-specific wind-wave directional patterns.

\section{ Discussion and Conclusion } \label{conclusion}
Several non-linear regression models through the M\"{o}bius transformation have been proposed in the literature to address relationships involving directional data. Notably, works by \cite{downs2002circular}, \cite{kato2008circular}, and more recently \cite{jha2017multiple, jha2018circular}, and \cite{biswas2025semi} have developed circular-circular and linear-circular regression models, primarily in contexts where a single circular predictor is linked with a circular or linear response. These models demonstrate the utility of
M\"{o}bius transformations in preserving angular structure and enabling effective modeling.

In a wider setup, our study extends this line of research by generalizing the M\"{o}bius transformation to accommodate two circular predictors and two circular responses. Thereby, we have proposed a new framework for torus-to-torus regression. This extension addresses considerably more complex modeling, particularly relevant for applications involving bivariate angular data. Despite the growing prevalence of such data in fields like meteorology, bioinformatics, and finance, there has been a noticeable gap in statistical methodology for modeling data with angular vectors. To the best of our knowledge, no prior work has established a regression framework operating entirely on the torus, $\mathbb{T}_2$.

The proposed model bridges the gap by offering a principled approach to torus-to-torus regression. We rigorously analyzed the identifiability and geometric structure of the model and introduced a novel loss function grounded in differential geometry. This loss function enables semi-parametric  estimation of the model parameters without the need to specify a particular distribution for the angular error, which increases the flexibility and adaptability with the data. Additionally, a new visualization method for circular data is presented here to facilitate a better understanding of both the observed and predicted data.

The practical applicability of our approach was demonstrated through its application to wind-wave direction observed during two major cyclonic events,  \textit{Amphan} and Biparjoy, that severely impacted the east and west coasts of India. These applications illustrate the capacity of the model to capture complex directional dependencies under dynamic environmental conditions regardless of the location and the nature of the cyclones and represent the first known application of torus-to-torus regression in a meteorological setting.

\bibliographystyle{unsrtnat}
\bibliography{buddha_bib}  






\end{document}